\documentclass[aps,prd,reprint,11pt, twocolumn, tightenlines, notitlepage, superscriptaddress, nofootinbib, preprintnumbers, floatfix]{revtex4-1}
\pdfoutput=1
\usepackage{textcomp}
\usepackage{epsfig,amsfonts,mathrsfs,amsmath,amssymb,graphicx,color,slashed,multirow}
\usepackage{amsmath,latexsym,amssymb,graphicx,slashed,hyperref,color,enumerate,url,etoolbox,cancel,gensymb}
\hypersetup{colorlinks,citecolor= nicegreen,linkcolor= nicered}
\definecolor{nicered}{rgb}{0.7,0.1,0.1}
\definecolor{nicegreen}{rgb}{0.1,0.5,0.1}

\usepackage{comment}

\usepackage[left=0.6in, right=0.6in, top=1in, bottom=0.6in]{geometry}

\definecolor{vdrgreen}{rgb}{0.0, 0.7, 0.0}

\definecolor{kjkblue}{rgb}{0.39, 0.589, 0.6914}

\def\Fermilab{Theoretical Physics Department, Fermilab, P.O. Box 500, Batavia, IL 60510, USA}
\def\Valencia{AHEP Group, Instituto de F\'{i}sica Corpuscular, CSIC/Universitat de Val\`{e}ncia, Calle Catedr\'{a}tico Jos\'{e} Beltr\'{a}n, 2 E-46980 Paterna, Spain}

\begin{document}
\title{Hunting On- and Off-Axis for Light Dark Matter with DUNE-PRISM}

\author{Valentina De Romeri}
\affiliation{\Valencia}
\author{Kevin J. Kelly}
\affiliation{\Fermilab}
\author{Pedro A.N. Machado}
\affiliation{\Fermilab}


\begin{abstract}
We explore the sensitivity of the Deep Underground Neutrino Experiment (DUNE) near detector and the proposed DUNE-PRISM movable near detector to sub-GeV dark matter, specifically scalar dark matter coupled to the Standard Model via a sub-GeV dark photon. We consider dark matter produced in the DUNE target that travels to the detector and scatters off electrons. By combining searches for dark matter at many off-axis positions with DUNE-PRISM, sensitivity to this scenario can be much stronger than when performing a measurement at one on-axis position.
\end{abstract}

\preprint{FERMILAB-PUB-19-116-T}

\maketitle 

\section{Introduction}\label{sec:Introduction}
Although dark matter (DM) is undoubtedly present in our universe, its detection via non-gravitational effects has eluded us~\cite{Aaboud:2017phn, Sirunyan:2017jix, Amole:2019fdf, Agnese:2018gze, Agnes:2018ves, Ren:2018gyx, Cui:2017nnn, Agnese:2017jvy, Aprile:2017iyp,
Akerib:2016vxi, Fermi-LAT:2016uux, Ahnen:2016qkx, Aguilar:2016kjl, Aartsen:2016zhm, Adrian-Martinez:2016gti,Bertone:2004pz}. One well-motivated hypothesis regarding DM is that, in the early universe, it was in thermal equilibrium with the standard model (SM) plasma before its interactions froze out, resulting in a relic abundance that persists today~\cite{Gondolo:1990dk}. One scenario that fits this description is that of a light dark sector where a DM particle interacts with the SM via a new gauge boson.

\begin{figure}
\begin{center}
\includegraphics[width=\linewidth]{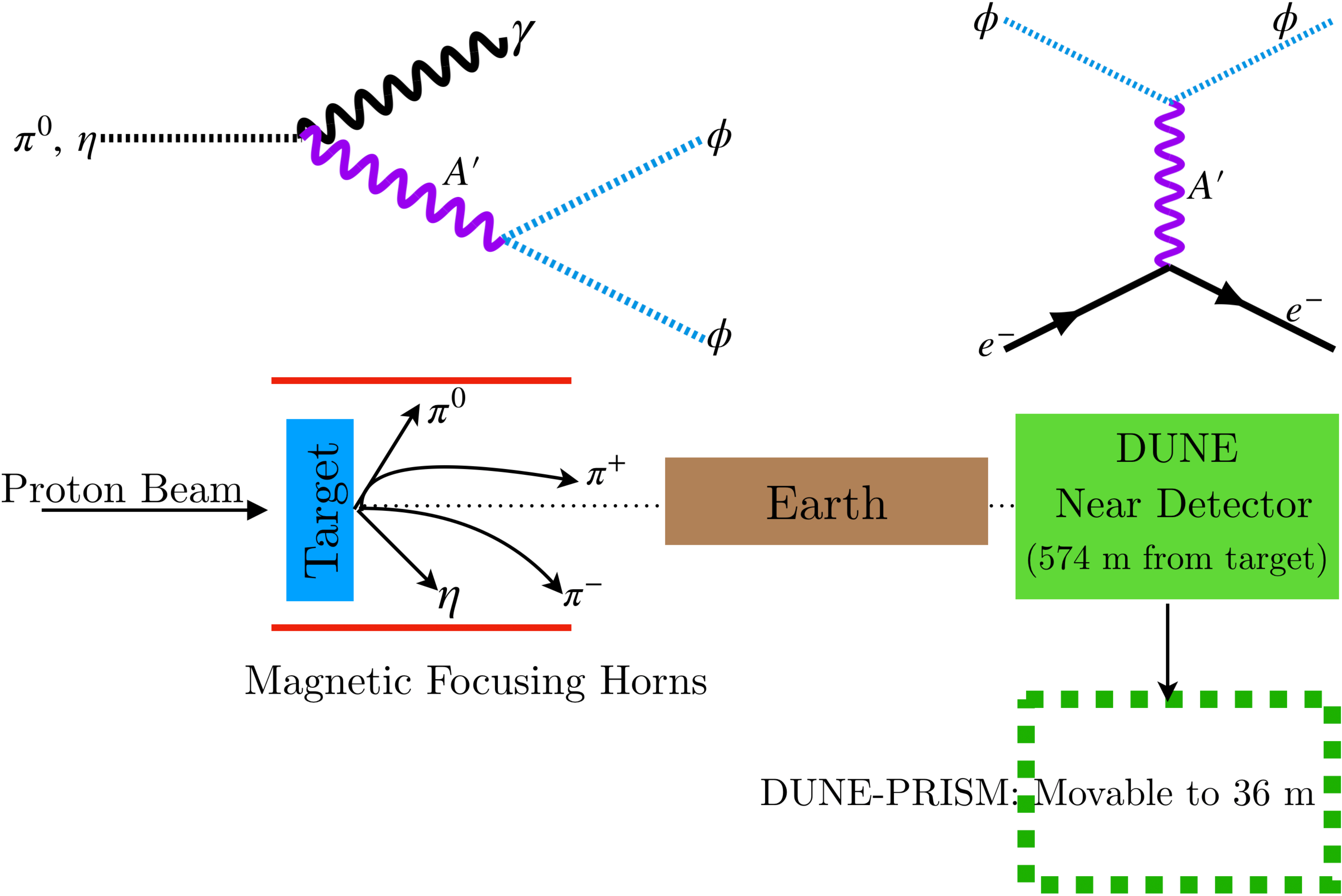}
\caption{Schematic setup of the proposed search for dark matter using DUNE-PRISM. This diagram is not to scale. See Refs.~\cite{BrossTalkPONDD,VilelaTalkPONDD} for more detailed schematics.\label{fig:ExpSetup}}
\end{center}
\end{figure}

Recently, significant attention has been paid to the prospects of detecting sub-GeV DM in neutrino detectors, leveraging the accompanying intense proton beams of these experiments~\cite{Batell:2009di, deNiverville:2011it, deNiverville:2012ij, Izaguirre:2013uxa, Coloma:2015pih, deNiverville:2016rqh, deNiverville:2018dbu, Jordan:2018gcd, deGouvea:2018cfv, Kelly:2019wow}. DM can be produced in the collision of protons on a target and travel to a near detector, interacting with nuclei or electrons -- Fig.~\ref{fig:ExpSetup} provides a schematic picture of this concept. Since DM interactions would look very similar to neutral current neutrino interactions, a usual way to reduce the neutrino background is to look at events off the beam axis~\cite{Coloma:2015pih,deGouvea:2018cfv}. Neutrinos come from charged meson decays, which are focused by a magnetic horn system in the forward direction, while DM is produced via the decay of neutral, unfocused mesons. Therefore, the signal-to-background ratio of DM to neutrino events grows for larger off-axis positions.

In this paper, we focus on the possibility of the future Deep Underground Neutrino Experiment (DUNE)~\cite{Abi:2018dnh} to probe such a DM scenario. Specifically, we focus on the proposed DUNE-PRISM concept~\cite{VilelaTalkPONDD} in which the near detector moves up to ${\sim}36$~m off-axis. We show that performing searches for DM at several off-axis locations provides a sensitivity much stronger than performing a search at any one location by reducing correlated uncertainties regarding the neutrino/DM flux and cross sections. Even with reduced statistics from moving off-axis, such a search can probe significantly more parameter space for the light dark matter scenario with the same amount of time collecting data.

This manuscript is organized as follows: in Section~\ref{sec:LDM}, we discuss how such light dark matter particles are produced in a neutrino facility. For clarity, we focus on scalar DM; fermionic DM is discussed in the Appendices. Section~\ref{sec:SignalsBackgrounds} discusses the signals (and their associated backgrounds) of interest for this study. We also discuss the advantages of having both on- and off-axis measurements concretely, and explain our statistical procedures for this search. Section~\ref{sec:ExistingLimsDUNE} discusses existing limits on this scenario and presents our results, and finally, Section~\ref{sec:Conclusions} offers some concluding remarks.

\section{Light Dark Matter in a Neutrino Facility}\label{sec:LDM}
The strength of neutrino facilities searching for sub-GeV DM comes from the production mechanism, in which protons strike a target, producing an abundance of charged and neutral mesons. These may decay to DM, which in turn travel to a detector producing a wide range of possible signals~\cite{Batell:2009di, deNiverville:2011it, deNiverville:2012ij, Izaguirre:2013uxa, Coloma:2015pih, deNiverville:2016rqh, deNiverville:2018dbu, Jordan:2018gcd, deGouvea:2018cfv,deGouvea:2018cfv}.

We focus on a $U(1)_D$ dark photon scenario in which the dark photon $A'$ mixes kinetically with the photon, and a new complex scalar $\phi$ is charged under $U(1)_D$. The Lagrangian of interest is
\begin{equation}
\mathcal{L} \supset -\frac{\varepsilon}{2} F^{\mu\nu} F^\prime_{\mu\nu} + \frac{M_{A^\prime}^2}{2} A_\mu^\prime A^{\prime \mu} + \left \lvert D_\mu \phi\right\rvert^2 - M_\phi^2 \left\lvert \phi\right\rvert^2,
\end{equation}
where $\varepsilon$ is the kinetic mixing between the SM and the new $U(1)_D$ (which has gauge coupling $g_D$ or dark fine structure constant $\alpha_D \equiv g_D^2/4\pi$), and $M_{A^\prime}$ and $M_\phi$ are the dark photon and DM masses. Such a scenario is appealing for its minimality; we only need to introduce two particles (with masses) and two couplings to explain the relic abundance of dark matter. An equally minimal scenario exists if the dark matter is fermionic, however, more stringent constraints apply in this scenario. We discuss the possibility of fermionic DM in the Appendices.

In this work, we assume that the DM is a thermal relic and that its initial abundance is symmetric. In this case, the DM and $A^\prime$ masses and couplings will provide a target for which the relic abundance matches the observed abundance in the universe. We discuss this target in Section~\ref{sec:ExistingLimsDUNE}, and will show that our projected DUNE-PRISM sensitivity will reach this target for some combinations of dark matter and $A^\prime$ masses.

\begin{figure}
\begin{center}
\includegraphics[width=\linewidth]{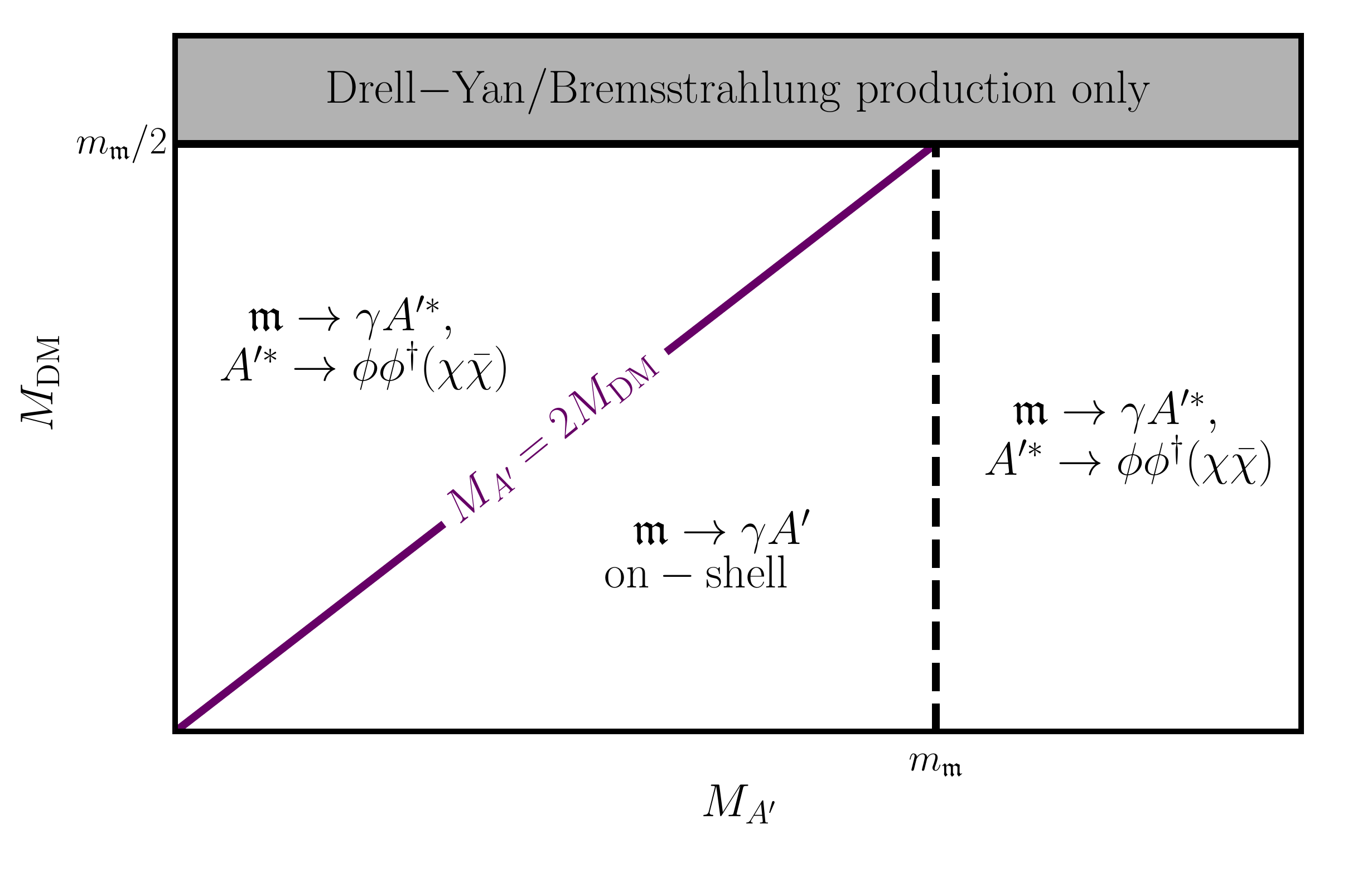}
\caption{Regions of parameter space for $A^\prime$ mass $M_{A^\prime}$ and dark matter mass $M_\mathrm{DM}$ (either fermionic or scalar) for which different meson decay production mechanisms exist. In the lower-triangular region, all decays proceed on shell, and we expect the most sensitivity in this region. In the grey shaded region, production via meson decay cannot occur; only production via direct Drell-Yan production or proton bremsstrahlung may contribute.\label{fig:PhaseSpaceBlank}}
\end{center}
\end{figure}

\textbf{Dark Matter Production:} We focus on the region of parameter space in which the dark matter mass $M_\phi$ is less than half the mass of a pseudoscalar meson $\mathfrak{m}$ which is produced in the DUNE target. DM is produced via two decays, those of on-shell $A^\prime$ and those of off-shell $A^\prime$ (if $M_{A^\prime} > m_\mathfrak{m}$ or $M_{A^\prime} < 2M_\phi$). The regions of parameter space for which these two processes occur is shown in Fig.~\ref{fig:PhaseSpaceBlank} -- on-shell decays proceed in the lower-triangular region.

We use {\sc Pythia8}~\cite{Sjostrand:2007gs} to estimate the production of $\pi^0$ and $\eta$ mesonsand find that, on average, for a 120~GeV proton beam, $4.5$ $\pi^0$ and $0.5$ $\eta$ are produced per proton-on-target (POT). The branching fractions (less than $10^{-6}$ in general) into DM are derived in the Appendix. Our simulated dark matter production consists of only the $\pi^0$ and $\eta$ produced in the primary interaction of protons on target. DM may also be produced via $K^\pm \to \pi^\pm \pi^0$ followed by $\pi^0 \to \gamma A^\prime$.  Kaon production is subdominant relative to pion production and the aforementioned branching ratio is about 20\%. Thus, to avoid the hurdle of simulating the magnetic horn focusing of charged particles, we neglect this contribution.

For any combination of $M_\phi$ and $M_{A^\prime}$, we can estimate the differential DM flux (as a function of energy) that reaches the DUNE near detector, allowing us to assess the experimental sensitivity to light DM scenarios. As indicated by the grey region in Fig.~\ref{fig:PhaseSpaceBlank}, if the dark matter mass is above $m_\mathfrak{m}/2$, then production can proceed via direct, Drell-Yan production, or proton bremsstrahlung via the process $p p \to p p A^\prime$, $A^\prime \to \phi \phi^\dagger$. Drell-Yan production is only significant for dark matter masses $\gtrsim 1$ GeV, where existing limits from other searches would dominate over such a search. We have simulated bremsstrahlung production for the experimental setup of interest and found that it only exceeds the meson production when $M_\phi > m_\eta/2$ -- see Refs.~\cite{Blumlein:2013cua,deNiverville:2016rqh} for a detailed description of the bremsstrahlung production cross section. Our sensitivity should then be viewed as a robust, conservative sensitivity estimate for the region in which we search -- a search using the bremsstrahlung contribution for heavier masses could add to the sensitivity for a narrow region of parameter space\footnote{As we will discuss in the following sections, a key advantage of our search strategy is that the dark matter flux is broad as one begins to move the detector off-axis. This is because it is produced from neutral meson decays, where the neutral mesons are unfocused. The proton bremsstrahlung process peaks at low transverse momentum, meaning it will drop off more sharply as one goes off-axis. We expect that, while a search using this production mechanism could exist, it would not be quite as powerful as the search we present here.}.

\section{Signals and Backgrounds at the DUNE Near Detector}\label{sec:SignalsBackgrounds}
DUNE is designed primarily to study neutrino oscillations and interactions, however the far and near  detectors will be powerful tools for searching for new physics. We assume that the near detector, with dimensions $3$ m $\times$ 4 m $\times$ 5 m and 75~t fiducial mass, is located 574 m downstream of the production target. For DM signals, we will focus\footnote{One may also consider scattering of the DM off the nucleus. This process has a larger cross section, leading to a significantly larger number of signal events. However, the corresponding background, neutrino neutral-current scattering, is also significantly larger, making such a search difficult to perform.} on scatterings off electrons in the liquid argon, DM $+~e^- \to$ DM $+~e^-$ (Fig.~\ref{fig:ExpSetup} right), the cross sections for which are given in the Appendix. Neutrinos will produce events that look similar to this, either via $\nu e^- \to \nu e^-$ or through charged-current quasi-elastic (CCQE) scattering $\nu_e n \to e^- p$, in which the proton is not identified.

Our signal DM $+~e^-\to$ DM $+~e^-$ will look very similar to the $\nu e^- \to \nu e^-$ background due to the forward-going nature of the electron. The CCQE background can be safely vetoed by an ordinary cut on the electron energy and angle, $E_e\theta_e^2$ (see Appendix~\ref{Appendix:Veto} for details). Background events from $\pi^0$ mis-identification are very small due to LArTPC topology reconstruction and will also be vetoed by the $E_e\theta_e^2$ cut. Finally, it has been shown that the cosmic muon backgrounds can be strongly suppressed via timing, topology and fiducial volume cuts, and thus we consider them negligible here~\cite{Adams:2019bzt}.

\subsection{Advantages of searching on- and off-axis}\label{subsec:OffAxis}
\begin{figure*}[ht]
\begin{center}
\includegraphics[width=0.8\linewidth]{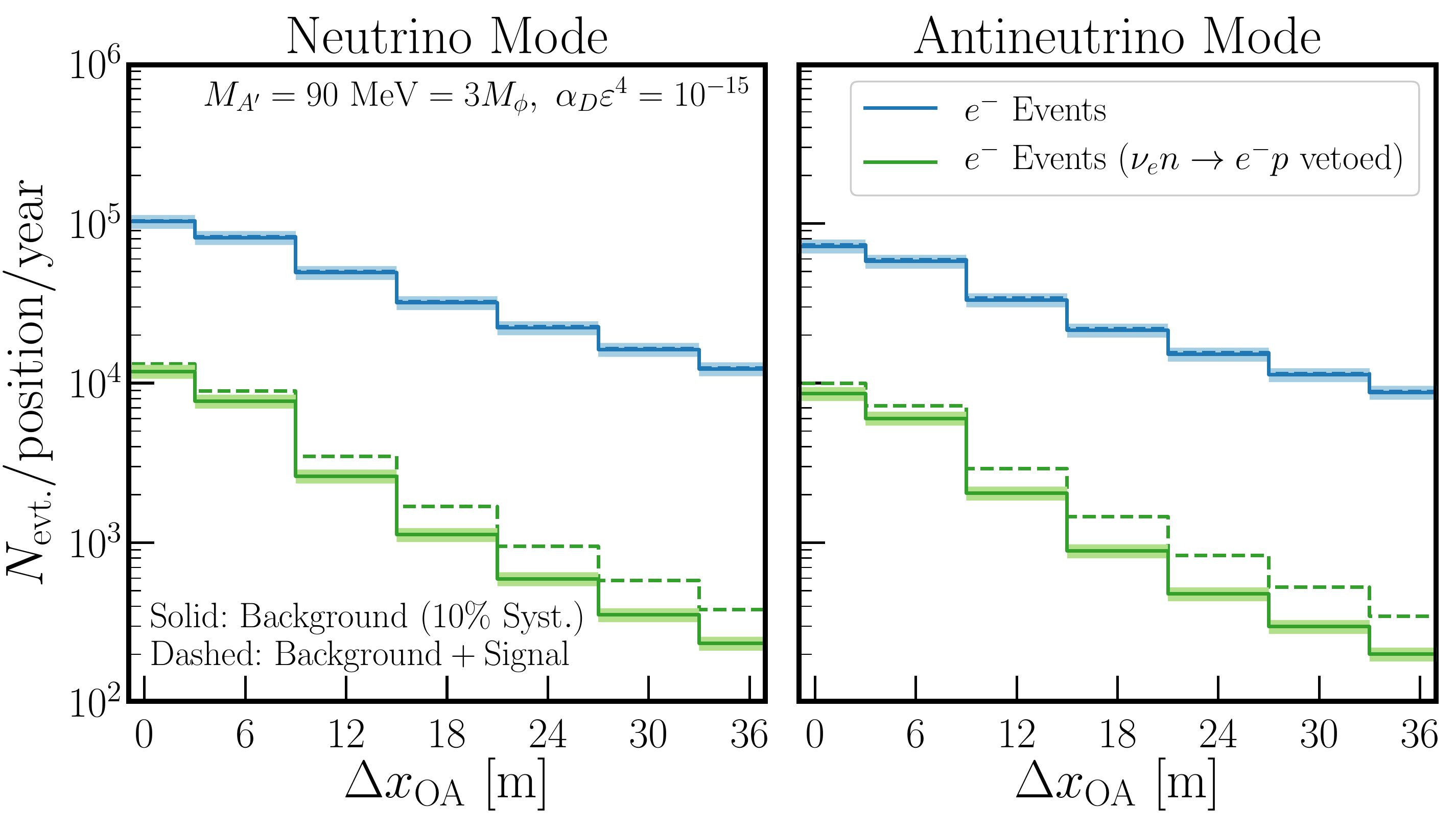}
\caption{Expected number of events per year in neutrino mode (left) and antineutrino mode (right) as a function of detector off-axis distance. We show event rates for electron scattering considering all backgrounds in blue, and with the CCQE background vetoed in green. We also display signal plus background events in dashed lines. Colored bands represent 10\% systematic uncertainty on background distributions.}
\label{fig:EvtsOffAxis}
\end{center}
\end{figure*}

With a movable near detector, the DUNE-PRISM concept will allow for precise measurements of the neutrino flux and cross sections~\cite{DUNE-PRISM,VilelaTalkPONDD}. 
For concreteness, we assume that the near detector will be able to move between $0$ and $36$ m transverse to the beam direction.
Neutrinos are produced via the decays of charged mesons which are focused by a magnetic horn system, while DM production occurs through decays of neutral mesons where no focusing is present.
Therefore, the neutrino-to-DM flux ratio decreases as one goes off-axis.
Fig.~\ref{fig:EvtsOffAxis} portrays this effect: we show the expected number of background (solid) and signal + background (dashed) events as a function of off-axis position, assuming one year of data collection (at each position), and $M_{A^\prime} = 90$ MeV, $M_\phi = 30$ MeV, $\alpha_D \varepsilon^4 = 10^{-15}$. We show these curves both without (blue) and with (green) the CCQE background veto. 
By comparing the blue and green curves, the importance of vetoing the CCQE background becomes evident.
Moreover, an increase in the signal-to-background ratio is clearly observed at further off-axis positions, particularly for $\Delta x_\mathrm{OA}\gtrsim12$~m.
Neutrino backgrounds were calculated according to the fluxes in Ref.~\cite{DUNEfluxes}.

In Fig.~\ref{fig:EvtsOffAxis} we display only a band of 10\% systematic uncertainty on the expected background events (we will discuss uncertainties in Section~\ref{subsec:Statistical}). Statistical error bars are too small to be visible on this scale.

\subsection{Statistical Tests and Systematic Uncertainties}\label{subsec:Statistical}
\begin{figure}[ht]
\begin{center}
\includegraphics[width=\linewidth]{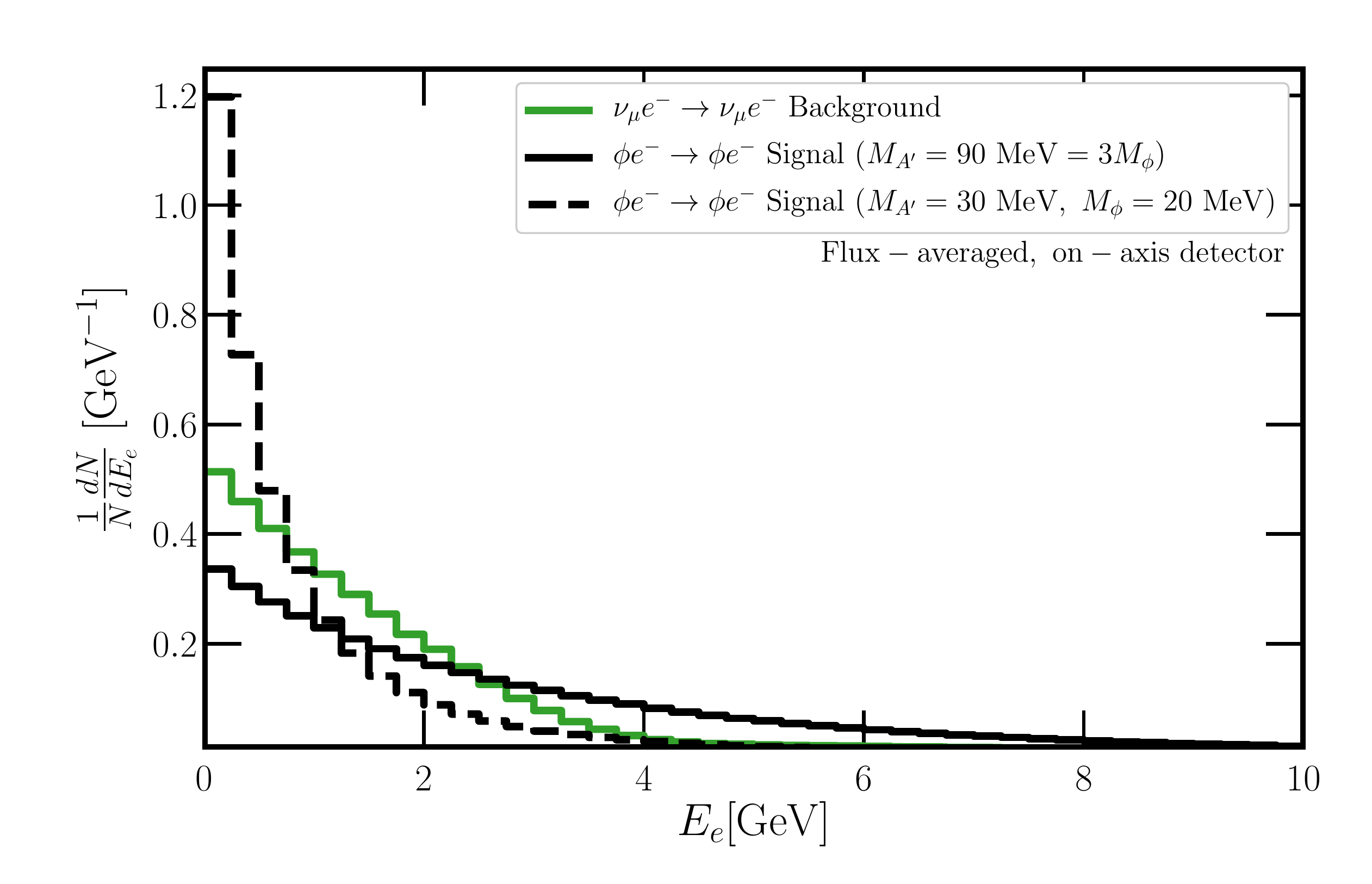}
\caption{Electron energy distribution for $\nu_\mu e^-$ background events (green) and DM $e^-$ scattering signal (black), with $M_{A^\prime} = 90$ MeV $= 3M_\phi$ (solid) and $M_{A^\prime} = 30$ MeV, $M_\phi = 20$ MeV (dashed) for an on-axis detector. Distributions are normalized.}
\label{fig:ElectronEnergyMainText}
\end{center}
\end{figure}

The main reason the DUNE-PRISM concept has been proposed is to reduce the systematic uncertainties associated with measuring the neutrino flux and cross section at the near detector. We leverage that fact in a DM search by combining several off-axis measurements. For simplicity, we assume that there is an overall flux-times-cross-section uncertainty $\sigma_A = 10\%$ correlated over all off-axis positions (but independent for neutrino and antineutrino beam modes). This is a correlated uncertainty because the relative fluxes at different off-axis positions are determined simply from the kinematics of meson decays. We assume additional independent normalization uncertainties $\sigma_{f_i} = 1\%$ at each off-axis position $i$. These uncertainties are included in our test statistic as nuisance parameters with Gaussian priors, and then marginalized over in producing a resulting sensitivity reach. We also include measurements of electron recoil energy in our analysis, by binning the expected signal and background distributions in broad $250$ MeV bins. Fig.~\ref{fig:ElectronEnergyMainText} displays a subset of the data in these distributions -- more detail on this procedure is given in Appendix~\ref{Appendix:Improvement}. The different shapes of these distributions provides additional power to search for DM.

Because the correlated flux uncertainty $\sigma_A$ is significantly larger than $\sigma_{f_i}$, we expect that sensitivity will be far greater for an analysis that divides data collection across several off-axis positions than one that collects all its data on-axis. For all of our analyses, we will assume $3.5$ years of data collection each in neutrino and antineutrino modes. We compare results assuming either all data is collected on-axis, or data collection is divided equally among all off-axis positions, $0.5$ yr at each position $i$. Appendix~\ref{Appendix:Improvement} details the test statistic and how we treat the nuisance parameters associated with the different uncertainties considered.

\section{Existing Limits and DUNE Sensitivity}\label{sec:ExistingLimsDUNE}
In this section, we discuss existing limits on sub-GeV dark matter for the region of parameter space of interest. In that context, we present the expected DUNE sensitivity and compare against existing and future searches for this type of dark matter.

\subsection{Theoretical Targets and Existing Constraints}
\paragraph{Theoretical constraints} 
In general, models with sub-GeV DM lead to an overabundance of relic particles~\cite{Lee:1977ua}, inspiring the inclusion of light mediators~\cite{Fayet:2004bw,Fayet:2006sp,Boehm:2003hm,Fayet:2007ua,Pospelov:2007mp}. Theoretical considerations of the running of the $U(1)_D$ fine structure constant limit $\alpha_D \lesssim 0.1-0.5$~\cite{Davoudiasl:2015hxa}.

\paragraph{Dark matter relic abundance}
The relationship between $M_{\rm DM}$ and $M_{A^\prime}$ plays an important role in the way the relic abundance freezes out thermally in the early Universe. 
The annihilation rate $\langle \sigma_\mathrm{ann} v\rangle$ primarily controls its present relic abundance, assuming no initial DM asymmetry, that is, identical number densities of $\phi$ and $\phi^\dagger$ (or $\chi$ and $\overline{\chi}$). 
The process that drives this annihilation is DM annihilation to $A^\prime A^\prime$ if  $M_{\rm DM} > M_{A^\prime}$, or to a pair of SM fermions via off-shell $A'$ if not~\footnote{
  In case the DM mass is below the $A'$ mass but not too much, the thermal DM distribution may still allow it to annihilate to $A'A'$, accounting for the correct relic abundance~\cite{DAgnolo:2015ujb,Cline:2017tka}. We do not investigate this scenario.}.
The tree-level annihilation cross section at relative velocity $v$ for a fermion dark matter candidate (see the appendices for more detail) is given by $s$-channel diagrams with $e^+ e^-$, $\mu^+ \mu^-$ and light hadrons in the final state~\footnote{The different thermal freeze-out channels have been studied in detail, see, e.g., Ref.~\cite{Pospelov:2007mp} for more detail.}: 
\begin{eqnarray}
\langle \sigma_\mathrm{ann} v \rangle_F &=& 24 \pi \alpha_D  \varepsilon^2  \frac{M_\chi }{ (4M_\chi^2 - M_{A^\prime}^2)^2} \Gamma(2 M_\chi) \nonumber \\
 &+&\Theta(M_\chi - M_{A^\prime}) \frac{\pi \alpha_D^2}{M_\chi^2} \frac{\left(1-\frac{M_{A^\prime}^2}{M_\chi^2}\right)^{3/2}}{\left(1-\frac{M_{A^\prime}^2}{2 M_\chi^2}\right)^2}, 
\label{eq:annxsec_F}
\end{eqnarray}
where $\Gamma$ is the dark photon width calculated at $M_{A^\prime} = 2 M_\chi$ and the second term enters if $M_{\chi} > M_{A^\prime}$ ($\Theta$ denotes the Heaviside theta-function).  For a scalar dark matter candidate, the annihilation cross section has a $v^2$ dependence which leads to a $p$-wave suppression of the annihilation rate at low velocities~\cite{Griest:1990kh}:
\begin{eqnarray}
\langle \sigma_\mathrm{ann} v \rangle_S &=& 4 \pi \alpha_D v^2 \varepsilon^2 \Gamma(2 M_\phi) \frac{M_\phi \left(1+\frac{m_f}{2 M_\phi}\right)}{ (4M_\phi^2 - M_{A^\prime}^2)^2} \nonumber \\
&+& \Theta(M_\phi - M_{A^\prime}) \frac{\pi \alpha_D^2}{M_\phi^2} \sqrt{1-\frac{M_{A^\prime}^2}{M_\phi^2}},
\label{eq:annxsec_S}
\end{eqnarray}

where $m_f$ stands for the mass of the SM fermion in the final state.
The requirement that $\phi~(\chi)$ comprises all the dark matter, $\Omega_{\phi (\chi)} = \Omega_\mathrm{obs} = 0.1186~h^{-2}$~\cite{Tanabashi:2018oca} demands $\langle \sigma_\mathrm{ann} v \rangle \sim 1$~pb. 
As we can see here, this only translate into a required value of $\varepsilon$ if $M_{\rm DM} < M_{A^\prime}$. 
Thermal relic abundance provides a target sensitivity for DUNE-PRISM, below which the annihilation cross section is too small, leading to a large DM abundance, incompatible with experimental observations. 
If we assume an initial DM asymmetry, an even larger annihilation rate would be necessary to deplete the symmetric DM component~\cite{Nussinov:1985xr,Kaplan:2009ag}. In this case, the thermal relic target is still useful as an experimental sensitivity goal.

\paragraph{Cosmic Microwave Background}
Precision measurements of the Cosmic Microwave Background (CMB) by the Planck satellite set a lower limit on the quantity $\Omega_{\phi (\chi)}^2 \times \langle \sigma_\mathrm{ann} v_\mathrm{CMB}\rangle$. 
If this quantity is too large, late-time annihilations of light dark matter can reionize Hydrogen and distort the CMB spectrum at high multipoles~\cite{Ade:2015xua,Slatyer:2015jla,Padmanabhan:2005es,Slatyer:2009yq,Finkbeiner:2011dx,Lin:2011gj,Galli:2011rz,Madhavacheril:2013cna}. 
This process depends on the annihilation cross section at the time of reionization. 
For masses of interest, the DM we are considering freezes out much before reionization, leading to much smaller  relative velocities of DM particles at reionization compared to freeze out. 
In the case of scalar dark matter, the annihilation cross section is $p$-wave suppressed~\cite{Boehm:2003hm}, that is, $\langle \sigma_\mathrm{ann} v\rangle \propto v^2$, and hence this limit becomes very weak.
This is also the case for Majorana or Pseudo-Dirac DM~\cite{TuckerSmith:2001hy}. 
If DM is fermionic and asymmetric, this limit still applies, but in a slightly different way, constraining the quantity $\Omega_{\chi} \Omega_{\overline{\chi}}  \times \langle \sigma_\mathrm{ann} v_\mathrm{CMB}\rangle$~\cite{Lin:2011gj}. We show results for fermionic DM in the Appendices. Specifically, the upper limit placed by Planck is shown in dashed orange in Fig.~\ref{fig:FermionLimit}, excluding the region below the line in each panel.

\paragraph{Beam-dump experiments}
A number of experiments similar in spirit to what we have proposed with DUNE-PRISM exist. Typically, dark photons are produced in a beam dump and (1) decay to dark matter which scatters off particles in the detector or (2)  propagate and decay to visible particles in a detector. 
The most stringent constraints using the first signature, for the parameter region of interest, comes from the Liquid Scintillator Neutrino Detector (LSND) experiment~\cite{deNiverville:2011it,Kahn:2014sra,Auerbach:2001wg,deNiverville:2018dbu},
the electron beam dump experiment E137~\cite{Bjorken:1988as, Batell:2014mga}, and the dedicated beam dump run of the MiniBooNE experiment~\cite{Aguilar-Arevalo:2018wea}. 
If $M_{A^\prime} < 2 M_{\rm DM}$, several existing and proposed experiments are sensitive to signature (2) where dark photons decay visibly, usually to $e^+ e^-$ or $\mu^+ \mu^-$~\cite{Pospelov:2008zw,Essig:2013lka}. Stringent constraints can be set using a vast array of searches at experiments such as Orsay~\cite{Davier:1989wz}, NA48/2~\cite{Batley:2015lha}, E137~\cite{Bjorken:1988as}, E141~\cite{Riordan:1987aw,Bjorken:2009mm}, E774~\cite{Bross:1989mp}, NA64~\cite{Banerjee:2017hhz,Gninenko:2019qiv}, muon or electron $g - 2$~\cite{Endo:2012hp,Davoudiasl:2015hxa}, KLOE~\cite{Babusci:2012cr} among others.


\paragraph{B-factories}
Mono-(dark)photon searches at BaBar provide stringent constraints on light DM scenarios via either of the processes $e^+ e^-  \rightarrow \gamma A^{\prime*} \rightarrow  \gamma~\textrm{invisible}$~\cite{Lees:2017lec} 
or $e^+ e^-  \rightarrow \gamma A^{\prime*} \rightarrow  \gamma~\ell^+ \ell^-$ , if $M_{A^\prime} < 2 M_{\rm DM}$~\cite{Lees:2014xha}.

\paragraph{Other cosmological and astrophysical observations}
For DM masses on the MeV scale, their freeze out may disturb big bang nucleosynthesis processes, regardless of whether the DM is a scalar or fermion. These constraints typically impose $M_{\rm DM} \gtrsim$ few MeV~\cite{Serpico:2004nm,Hisano:2008ti,Hisano:2009rc,Henning:2012rm,Knapen:2017xzo}.
Moreover, a coupling between sub-GeV dark photons and dark matter also give rise to   dark matter self-interactions. This may affect structure formation, DM halo distributions~\cite{MiraldaEscude:2000qt,Vogelsberger:2012ku}, and observations of the galaxy cluster collisions, such as the Bullet Cluster~\cite{Spergel:1999mh,Markevitch:2003at}. These measurements constrain the DM-DM scattering cross section to be  $\sigma_{\mathrm{DM}}/M_{\rm DM}<\mathcal{O}(1~{\rm barn}/{\rm GeV})$ for $M_{A^\prime} \lesssim 10$~MeV~\cite{Izaguirre:2013uxa,Izaguirre:2015yja}.

\paragraph{Direct Detection searches for Electron Scattering}
Direct detection experiments that traditionally search for dark matter-nucleon scattering, like XENON10~\cite{Essig:2012yx}/XENON100~\cite{Essig:2017kqs}, are sensitive to DM in this mass range via electron scattering. Such constraints require that the DM constitutes the entire relic abundance. For symmetric thermal relics and the parameters of interest, such DM would be depleted today, and these constraints do not apply. If considering an asymmetric DM scenario, these experiments are slightly more powerful than LSND.

\subsection{DUNE-PRISM Sensitivity}

\begin{figure*}[!htbp]
\begin{center}
\includegraphics[width=0.85\textwidth]{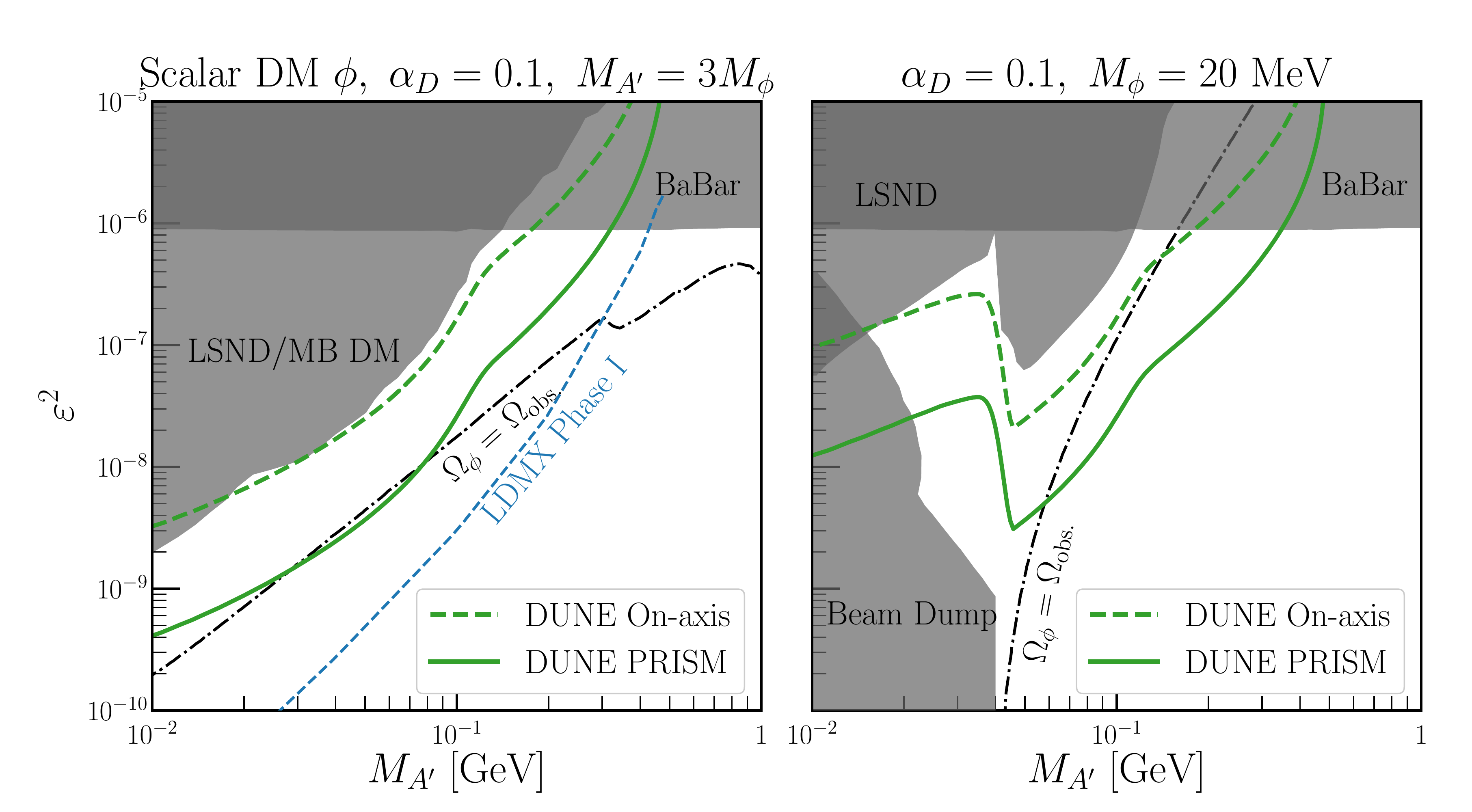}
\caption{Expected DUNE On-axis (dashed) and PRISM (solid) sensitivity at 90\% C.L. using $\phi e^- \to \phi e^-$ scattering. We assume $\alpha_D = 0.1$ in both panels, and $M_{A^\prime} = 3M_\phi$ ($M_\phi = 20$ MeV) in the left (right) panel. Existing constraints are shown in grey, and the relic density target is shown in a black dot-dashed line. We compare our results against the proposed LDMX experiment in blue~\cite{Akesson:2018vlm}.}
\label{fig:NSES}
\end{center}
\end{figure*}

We display expected DUNE 90\% confidence level sensitivity (including marginalization over the systematic uncertainty nuisance parameters discussed above and in Appendix~\ref{Appendix:Improvement}) assuming all on-axis data collection (DUNE On-axis) or equal times at each off-axis position (DUNE-PRISM) in Fig.~\ref{fig:NSES}, assuming\footnote{To present the most conservative results for this setup, we take the largest allowed value of $\alpha_D$ -- smaller values for $\alpha_D$ cause the observed relic abundance to appear easier to reach~\cite{Aguilar-Arevalo:2018wea}. Constraints on self-interacting dark matter in this mass range suggest $\alpha_D \lesssim 0.1$~\cite{Harvey:2015hha}.} $\alpha_D = 0.1$ and $M_{A^\prime} = 3M_\phi$ (left) or $M_\phi = 20$ MeV (right). The resulting sensitivity for fermionic DM is largely similar and shown in the Appendix. From our estimates, we see that DUNE can significantly improve the constraints from LSND~\cite{deNiverville:2018dbu} and the MiniBooNE-DM search~\cite{Aguilar-Arevalo:2018wea}, as well as BaBar~\cite{Lees:2017lec} if $M_{A^\prime} \lesssim 200$ MeV. We also show limits in the right panel from beam-dump experiments~\cite{Davier:1989wz,Batley:2015lha,Bjorken:1988as,Riordan:1987aw,Bjorken:2009mm,Bross:1989mp}, as well as the lower limit obtained from matching the thermal relic abundance of $\phi$ with the observed one (black, dot-dashed). In the right panel, we do not show the thermal relic abundance line below $M_{A^\prime} = 2M_\phi$; for $M_{A^\prime} < M_\phi$, freeze-out occurs via the process $\phi \phi^\dagger \to A^\prime A^\prime$, independent of $\varepsilon^2$. For $M_\phi < M_{A^\prime} < 2M_{\phi}$, this process still contributes to freeze-out due to non-zero temperature effects~\cite{DAgnolo:2015ujb,Cline:2017tka}, and a thermal relic target would require much smaller $\varepsilon^2$ than what we display here.

In the left panel of Fig.~\ref{fig:NSES}, we also display the expected sensitivity to this type of dark matter proposed by the LDMX experiment~\cite{Akesson:2018vlm}. We see that, while our proposed sensitivity is not as powerful as LDMX, using DUNE-PRISM to search for light dark matter is complementary to the proposal to use DUNE-PRISM to reduce systematic uncertainties in the study of neutrino oscillations, where LDMX is a dedicated experiment to search for light dark matter. Because these two sensitivities come from very different experiments, we caution the reader from making comparisons.

The BaBar experiment, which provides the most stringent constraints for $A^\prime$ masses above roughly 100 MeV, will be succeeded by the Belle II experiment. Belle II will be sensitive to invisibly decaying dark photons with a kinetic mixing of roughly $\varepsilon^2 \approx 10^{-7}$ by searching for the process $e^+ e^- \to \gamma A^\prime,$ $A^\prime \to \chi \overline{\chi}$~\cite{Kou:2018nap}.

\begin{figure}[tbp]
\begin{center}
\includegraphics[width=\linewidth]{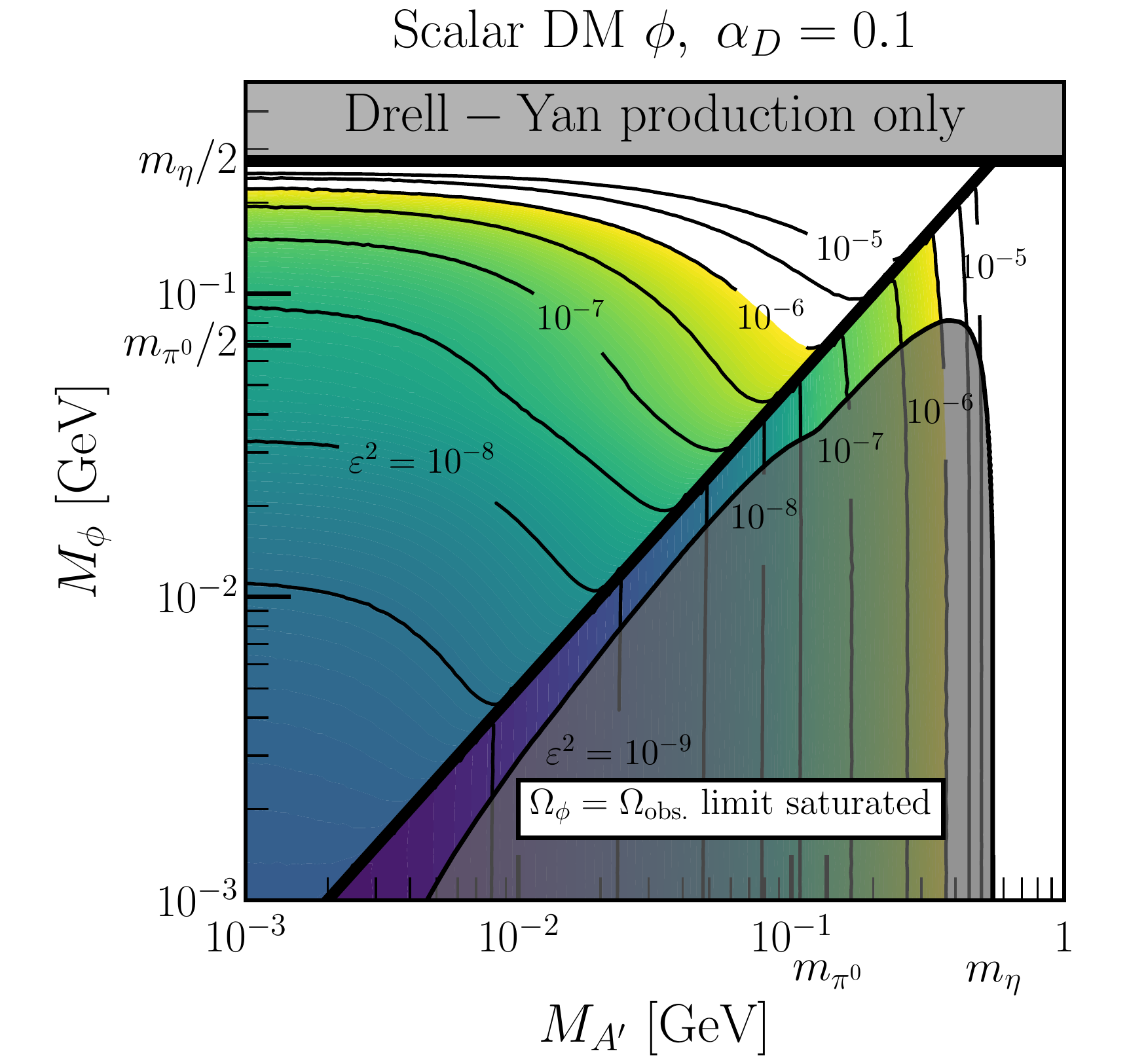}
\caption{Expected limits on $\varepsilon^2$ as a function of $M_{A^\prime}$ and scalar DM mass assuming seven years of data collection at DUNE searching for DM scattering off electrons (more detail in text). We shade the region for which this expected limit saturates the target for which the DM relic density matches the observed abundance.}
\label{fig:2DES}
\end{center}
\end{figure}

We also estimate limits on $\varepsilon^2$ varying $M_{A^\prime}$ and $M_\phi$ independently\footnote{Previously, we calculated limits by setting our $\Delta \chi^2$ function to be equal to $4.61$ for a 90\% CL estimate for two parameters. Here, we set it to $6.25$ for three.} in Fig.~\ref{fig:2DES}. Here, we also determine the values of $M_{A^\prime}$ and $M_{\phi}$ for which our limit on $\varepsilon^2$ reaches the thermal relic abundance target for $\phi$ (assuming it is a symmetric thermal relic). This region is colored in grey.
As discussed above cf the right panel of Fig.~\ref{fig:NSES}, we may not saturate either of these limits in the region $M_{A^\prime} < 2M_\phi$~\cite{DAgnolo:2015ujb,Cline:2017tka}. We repeat the procedure of Fig.~\ref{fig:2DES} assuming fermionic DM in the Appendix and in Fig.~\ref{fig:Fermion2D}. For fermionic DM, we reach the relic target for a smaller range of parameters, and additionally constraints from the Planck satellite~\cite{Ade:2015xua} are relevant.

\section{Discussion \& Conclusions}\label{sec:Conclusions}
In this paper we have estimated the sensitivity of the future DUNE experiment to light dark matter models taking into account the potential of the DUNE-PRISM detector. 
Two scenarios were considered for the estimate: scalar and fermionic dark matter below the GeV scale which interacts with the SM particles via a light dark photon kinetically mixed with the photon.
We have found that, in both cases, the experimental sensitivity is substantially increased by the DUNE-PRISM ability to look at events off the beam axis. An analysis with DUNE-PRISM will allow sensitivity to reach regions of parameter space predicted by simple, thermal relic dark matter models -- this will not be possible without a moving near detector.

In this way, DUNE-PRISM will be competitive with dedicated experiments in probing light dark matter scenarios. Specifically, we find that DUNE-PRISM will be sensitive to values of $\varepsilon^2$ only a factor of $\sim3$\footnote{Assuming $\alpha_D = 0.1$ and at $M_{A^\prime} = 3M_\phi = 90$ MeV. At smaller DM masses (or larger $\alpha_D$) this factor could get slightly worse, but not larger than $\sim 7$.} higher than those probed by phase I of LDMX, an experiment designed specifically to search for light dark matter~\cite{Akesson:2018vlm}. In this work we have shown that DUNE-PRISM, an experiment that is very likely to occur due to other scientific goals, will have competitive sensitivity to future, dedicated experiments. This fact is non-trivial and had not been previously shown in the literature.

\section*{Acknowledgments}
We are extremely grateful to Laura Fields for generating neutrino fluxes at off-axis locations. We thank Paddy Fox, Roni Harnik, Gordan Krnjaic, and Yue Zhang for useful discussions regarding this work, and Alberto Marchionni for encouragement. VDR is also very grateful for the kind hospitality received at Fermilab where this work was initiated.

This manuscript has been authored by Fermi Research Alliance, LLC under Contract No. DE-AC02-07CH11359 with the U.S. Department of Energy, Office of Science, Office of High Energy Physics.
VDR acknowledges financial support by the Fermilab Neutrino Physics Center Fellowship Award (Spring 2018) and by the ``Juan de la Cierva Incorporacion'' program (IJCI-2016-27736) funded by the Spanish MINECO, as well as partial support by the Spanish grants SEV-2014-0398, FPA2017-85216-P (AEI/FEDER, UE), PROMETEO/2018/165 (Generalitat Valenciana) and the Spanish Red Consolider MultiDark FPA2017-90566-REDC.

\appendix\setcounter{footnote}{0}
\begin{widetext}
\section{Derivation of Meson Decay Branching Fractions}
\label{Appendix:Branching}

In this section, we derive the expressions for the branching fraction of pseudoscalar mesons into both scalar and fermionic DM. The relevant Feynman diagrams for these decays are shown in Fig.~\ref{fig:MesonDecays}. We characterize the coupling between the pseudoscalar meson $\mathfrak{m}$ and two photons (or one photon and one dark photon via kinetic mixing) in terms of a dimension-five effective operator $-\frac{1}{4} A_{\mathfrak{m}\gamma\gamma} \mathfrak{m} F^{\mu\nu} \widetilde{F}_{\mu\nu}$. This will allow us to express our result in terms of the branching fraction into two photons, which is
\begin{equation}
\mathrm{Br}(\mathfrak{m} \to \gamma\gamma) = \frac{A_{\mathfrak{m}\gamma\gamma}^2 m_\mathfrak{m}^3}{64\pi}.
\end{equation}
\begin{figure}[!hbtp]
    \centering
    \includegraphics[width=0.30\linewidth]{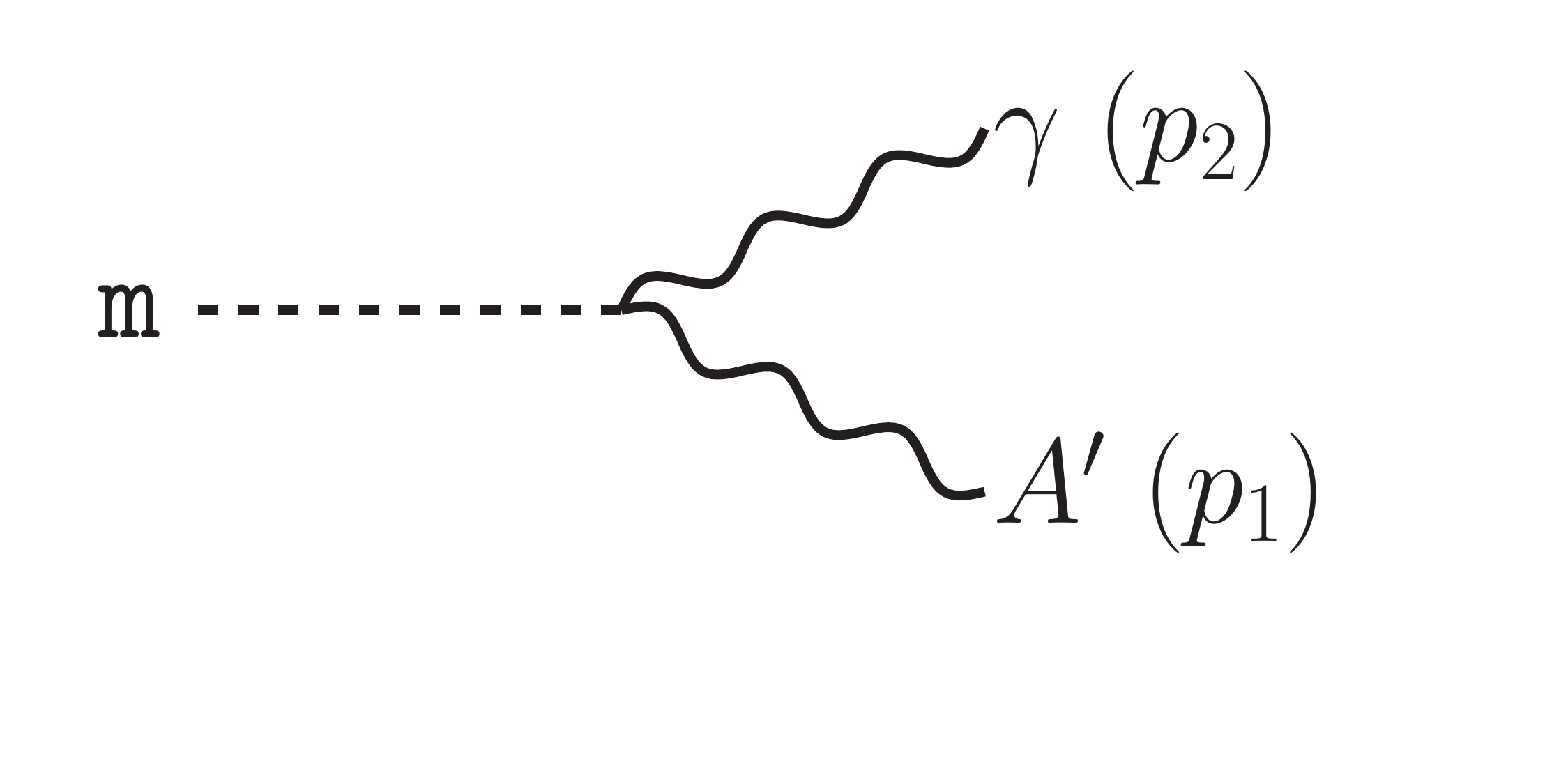}
    \includegraphics[width=0.30\linewidth]{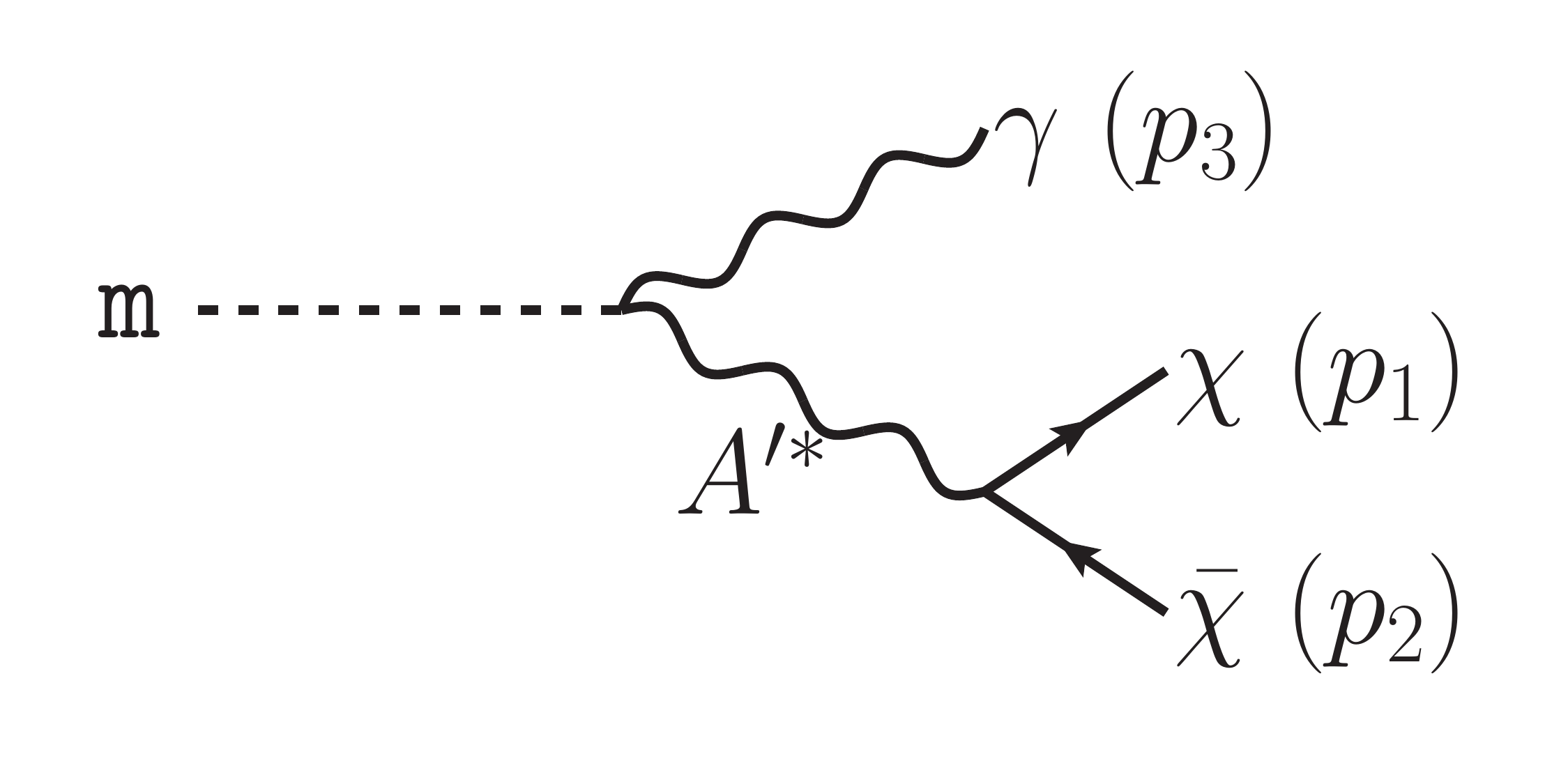}
    \includegraphics[width=0.30\linewidth]{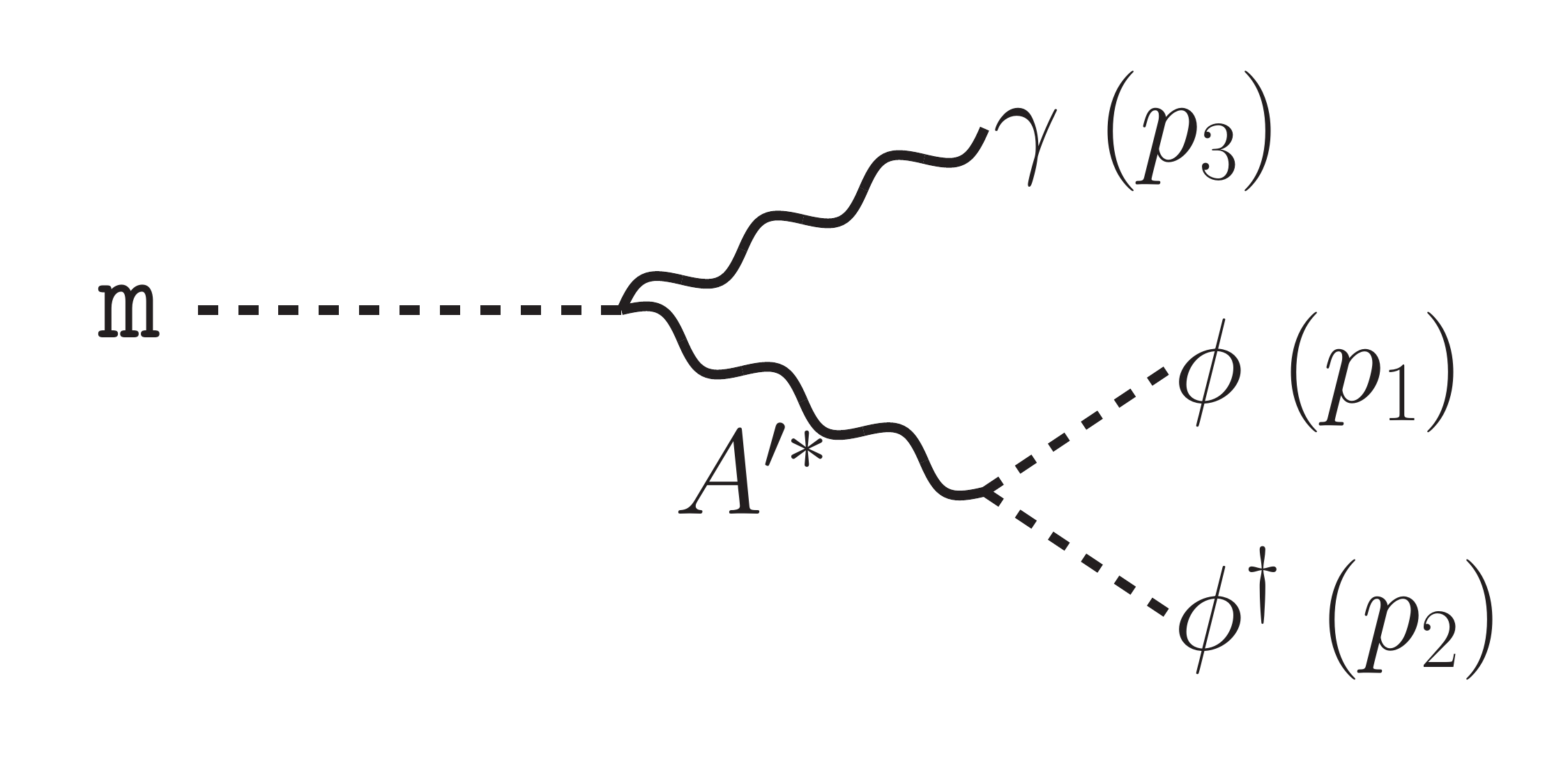}
    \caption{Meson decays of interest for this work. Left: two-body decay $\mathfrak{m} \to \gamma A^\prime$, when $M_{A^\prime} < m_\mathfrak{m}$. Center: three-body decay $\mathfrak{m} \to \gamma \chi \overline{\chi}$ for fermionic DM. Right: three-body decay $\mathfrak{m} \to \gamma \phi \phi^\dagger$ for scalar DM.}
    \label{fig:MesonDecays}
\end{figure}

When $2 m_{\rm DM} < M_{A^\prime} < m_\mathfrak{m}$ (first diagram in Fig.~\ref{fig:MesonDecays}) the dark photon is produced on-shell and we can use the narrow-width approximation. In this case, a neutral pseudoscalar meson will have a branching fraction of $\mathfrak{m} \to \gamma A^\prime$ proportional to its $\gamma\gamma$ branching ratio: 

\begin{eqnarray}
\mathrm{Br}(\mathfrak{m} \to \gamma \phi \phi^\dagger) = 
\mathrm{Br}(\mathfrak{m} \to\gamma \gamma) \times 2\varepsilon^2 \left(1 - \frac{M_{A^\prime}^2}{m_\mathfrak{m}^2}\right)^3 \times \mathrm{Br}(A^\prime \to \phi \phi^\dagger)~~~~~\rm(scalar~ DM, on -shell),
\end{eqnarray}

\begin{eqnarray}
\mathrm{Br}(\mathfrak{m} \to \gamma \chi\bar{\chi}) = 
\mathrm{Br}(\mathfrak{m} \to\gamma \gamma) \times 2\varepsilon^2 \left(1 - \frac{M_{A^\prime}^2}{m_\mathfrak{m}^2}\right)^3 \times \mathrm{Br}(A^\prime \to \chi\bar{\chi})~~~~~\rm(fermionic~DM, on -shell).
\end{eqnarray}

For the entirety of this work, as long as $M_{A^\prime} > 2 M_\mathrm{DM}$, we assume $\mathrm{Br}(A^\prime \to \phi\phi^\dagger (\chi\bar{\chi} )) = 1$ -- this assumption is correct as long as $\alpha_D \gg \varepsilon^2$ so that  decays of $A^\prime \to \ell^+ \ell^-$ are suppressed.

If  $ M_{A^\prime} < 2 M_{\rm DM}$ (or $ M_{A^\prime} > m_\mathfrak{m}$) the narrow-width approximation cannot be used and the DM is produced via a neutral meson three-body decay (second and third diagrams in Fig.~\ref{fig:MesonDecays}). In the following, we derive the expressions for the three-body decay branching fraction of $\mathfrak{m} \to \gamma\phi \phi^\dagger$ ($ \gamma \chi\bar{\chi}$) for scalar (fermionic) DM.

We label the final-state momenta $p_1$ for $\phi$ ($\chi$), $p_2$ for $\phi^\dagger$ (or $\overline{\chi}$), and $p_3$ for $\gamma$. We define invariants as $s_{ij} \equiv \left(p_i + p_j\right)^2$ and will use the relation $s_{12} + s_{13} + s_{23} = m_\mathfrak{m} + m_1^2 + m_2^2 + m_3^2 = m_\mathfrak{m}^2 + 2M_\chi^2$ to eliminate $s_{13}$ from our calculation. The matrix element for the diagram in Fig.~\ref{fig:MesonDecays} (right) is then

\begin{equation}
i\mathcal{M}^\mathrm{(scalar)} = \frac{A_{\mathfrak{m}\gamma\gamma} \varepsilon g_D}{\left(s_{12} - M_{A^\prime}^2 \right)} \epsilon_{\mu\nu\alpha\beta} p_3^\mu \left(p_1+p_2\right)^\alpha \left(p_1 - p_2\right)^\beta \varepsilon^*(p_3)^\nu,
\end{equation}

where $\epsilon_{\mu\nu\alpha\beta}$ is the Levi-Civita tensor and $\varepsilon^*(p_3)^\nu$ represents the outgoing polarization vector of the photon. For fermion $\chi$, the matrix element is
\begin{equation}
i\mathcal{M}^\mathrm{(fermion)} = \frac{A_{\mathfrak{m}\gamma\gamma} \varepsilon g_D}{\left(s_{12} - M_{A^\prime}^2 \right)} \epsilon_{\mu\nu\alpha\beta} p_3^\mu \left(p_1+p_2\right)^\alpha \bar{u}_{p_1} \gamma^\beta v_{p_2} \varepsilon^*(p_3)^\nu.
\end{equation}

After taking the matrix-elements squared and replacing dot products with $s_{ij}$, we arrive at

\begin{equation}
\left\lvert\mathcal{M}^\mathrm{(scalar)}\right\rvert^2 = \frac{|A_{\mathfrak{m}\gamma\gamma}|^2 \varepsilon^2 g_D^2}{\left(s_{12} - M_{A^\prime}^2\right)^2} \left[-m_\mathfrak{m}^4 M_\phi^2 + m_\mathfrak{m}^2 s_{12} (M_\phi^2 + s_{23}) - s_{12}\left(M_\phi^4 - 2M_\chi^2 s_{23} + s_{23}(s_{12} + s_{23})\right)\right].
\end{equation}

For fermion $\chi$, the matrix-element squared is
\begin{align}
\left\lvert\mathcal{M}^\mathrm{(fermion)}\right\rvert^2 &= \frac{|A_{\mathfrak{m}\gamma\gamma}|^2 \varepsilon^2 g_D^2}{\left(s_{12} - M_{A^\prime}^2\right)^2} \times \nonumber \\
&\left[ m_\mathfrak{m}^4 (2M_\chi^2 + s_{12}) - 2m_\mathfrak{m}^2 s_{12} \left(M_\chi^2 + s_{12} + s_{23}\right) 
+ s_{12}\left(2(M_\chi^4 - 2M_\chi^2 s_{23} + s_{23}(s_{12} + s_{23})\right) + s_{12}^2  \right].
\end{align}

Since these decays are isotropic in the $\mathfrak{m}$ rest frame, we may use the convention from Ref.~\cite{Tanabashi:2018oca}, 
\begin{equation}
\Gamma_{\mathfrak{m}\to\gamma \chi\bar{\chi}} = \frac{1}{\left(2\pi\right)^3} \frac{1}{32 m_\mathfrak{m}^3} \int \int \left\lvert \mathcal{M_{\mathfrak{m}\to\gamma \chi\bar{\chi}}}\right\rvert^2 ds_{23} ds_{12},
\end{equation}
where the integration limits of $s_{23}$ depend on the Dalitz plot,
\begin{equation}
\left(s_{23}\right)^\mathrm{max}_\mathrm{min} = \left(E_2^* + E_3^*\right) - \left( \sqrt{E_2^{*2} - M_\chi^2} \mp E_3^*\right),
\end{equation}
where $E_2^* = \sqrt{s_{12}}/2$ and $E_3^* = (m_\mathfrak{m}^2 - s_{12})/(2\sqrt{s_{12}})$. The limits on $s_{12}$ are between $4M_\chi^2$ and $m_\mathfrak{m}^2$.

After the integral over $s_{23}$, we have
\begin{equation}
\Gamma^\mathrm{(fermion)}_{\mathfrak{m}\to\gamma \chi\overline{\chi}} = \frac{|A_{\mathfrak{m}\gamma\gamma}|^2 \varepsilon^2 g_D^2}{256\pi^3 m_\mathfrak{m}^3} \int_{4M_\chi^2}^{m_\mathfrak{m}^2} \frac{2(m_\mathfrak{m}^2 - s_{12})^3 \sqrt{s_{12} - 4M_\chi^2} (2M_\chi^2 + s_{12})}{3\sqrt{s_{12}} \left(s_{12} - M_{A^\prime}^2\right)^2} ds_{12},
\end{equation}
and
\begin{equation}
\Gamma^\mathrm{(scalar)}_{\mathfrak{m}\to\gamma \phi\phi^\dagger} = \frac{|A_{\mathfrak{m}\gamma\gamma}|^2 \varepsilon^2 g_D^2}{256\pi^3 m_\mathfrak{m}^3} \int_{4M_\phi^2}^{m_\mathfrak{m}^2} \frac{(m_\mathfrak{m}^2 - s_{12})^3 (s_{12} - 4M_\phi^2)^{3/2}}{6\sqrt{s_{12}} \left(s_{12} - M_{A^\prime}^2\right)^2} ds_{12}.
\end{equation}

We substitute $x \equiv M_\chi^2/m_\mathfrak{m}^2$ ($x \equiv M_\phi^2/m_\mathfrak{m}^2$), $y \equiv M_{A^\prime}^2/m_\mathfrak{m}^2$, and $z \equiv s_{12}/m_\mathfrak{m}^2$, arriving at
\begin{equation}\label{eq:FermionAppendix}
\frac{\mathrm{Br}(\mathfrak{m} \to \gamma \chi \overline{\chi})}{\mathrm{Br}(\mathfrak{m}\to \gamma\gamma)} = \alpha_D \varepsilon^2 \times \frac{2}{3\pi} \int_{4x}^1 \frac{(1-z)^3 (2x+z)\sqrt{1-\frac{4x}{z}}}{(z-y)^2} dz,
\end{equation}
and
\begin{equation}
\frac{\mathrm{Br}(\mathfrak{m} \to \gamma \phi \phi^\dagger)}{\mathrm{Br}(\mathfrak{m}\to \gamma\gamma)} = \alpha_D \varepsilon^2 \times \frac{1}{6\pi} \int_{4x}^1 \frac{(1-z)^3 z\left(1-\frac{4x}{z}\right)^{3/2}}{(z-y)^2} dz.
\end{equation}

One can verify Eq.~(\ref{eq:FermionAppendix}) by calculating this ratio for $y = 0$, $x = m_e^2/m_{\pi^0}^2$, $\varepsilon = 1$, and $\alpha_D = \alpha_\mathrm{EM}$, arriving at $\mathrm{Br}(\pi^0 \to \gamma e^+ e^-) \simeq 0.0119 \mathrm{Br}(\pi^0 \to \gamma\gamma)$, in agreement with the observed $\mathrm{Br}(\pi^0 \to\gamma\gamma) = 0.98823$ and $\mathrm{Br}(\pi^0\to\gamma e^+ e^-) = 0.01174$~\cite{Tanabashi:2018oca}.

\section{Dark Matter Scattering Cross Sections}
\label{Appendix:CrossSections}
Here we explicitly list the differential and total cross sections we consider for dark matter scattering off electrons in the detector for the sake of calculating the expected signal distributions.\\

The differential cross sections, in terms of the recoiling kinetic energy of the electron, are~\cite{deNiverville:2011it}
\begin{equation}
\frac{d\sigma(\phi e^-\to\phi e^-)}{dE_\mathrm{rec.}} = \frac{4\pi \varepsilon^2 \alpha_D \alpha_\mathrm{EM}}{E_\phi^2 - M_\phi^2} \frac{2E_\phi m_e \left(E_\phi - E_\mathrm{rec.}\right) - M_\phi^2 E_\mathrm{rec.}}{\left(2 m_e E_\mathrm{rec.} + M_{A^\prime}^2\right)^2}, \quad\quad\quad\quad\quad\quad\quad\quad\quad\quad\quad\quad \mathrm{(Scalar\ DM)}
\end{equation}

\begin{equation}
\frac{d\sigma(\chi e^-\to\chi e^-)}{dE_\mathrm{rec.}} = \frac{4\pi \varepsilon^2 \alpha_D \alpha_\mathrm{EM}}{E_\chi^2 - M_\chi^2} \frac{2E_\chi^2 m_e - (E_\mathrm{rec.}-m_e)(M_\chi^2 + 2E_\chi m_e + 2 m_e^2 - E_\mathrm{rec.} m_e)}{\left(2E_\mathrm{rec.} m_e + M_{A^\prime}^2 - 2m_e^2\right)^2}. \quad\quad \mathrm{(Fermionic\ DM)}
\end{equation}

To obtain a total cross section, we change variables $Q^2 \to 2m_e E_\mathrm{rec.}$, where $E_\mathrm{rec.}$ is the recoiling kinetic energy of the electron, integrated between $E_\mathrm{rec.}^\mathrm{min.}$ for experimental detection and
\begin{equation}\label{eq:Ermax}
E_\mathrm{rec.}^\mathrm{max.} = \frac{2m_e \left(E_\phi^2 - M_\phi^2\right)}{M_\phi^2 + 2 m_e E_\phi + m_e^2}.
\end{equation}

\section{Background Reduction for Electron Scattering}
\label{Appendix:Veto}
In the main text we discussed the signal channel of electron scattering, DM $+~e^- \to$ DM $+~e^-$ and its associated backgrounds. We showed that, if performing solely a counting experiment, the largest background is from electron neutrino beam contamination with CCQE scattering, $\nu_e n \to e^- p$ or $\overline{\nu}_e p \to e^+ n$, where the final-state hadronic system is unidentified. In displaying sensitivities, we showed two scenarios, one in which this background exists, and one in which it is completely removed due to kinematical constraints. In this appendix, we explain the strategy by which this background is removed, leaving only the ``irreducible'' $\nu e^- \to \nu e^-$ background. 

We first restrict ourselves to the scenario of elastic scattering of any particle off an electron at rest in the lab frame. Here, kinematics restrict the outgoing energy and angle of the final-state electron, $E_e \theta_e^2 < 2 m_e$, where $E_e$ is the total energy of the electron and $\theta_e$ is the angle with respect to the incoming particle direction. DUNE near detector is expected to measure the electron energy and angle with a precision of $5\sim10\%$ and $1^\circ$, respectively~\cite{Acciarri:2015uup}.
For the $\nu e^- \to \nu e^-$ background and our signal DM $+~e^- \to$ DM $+~e^-$, this restriction will hold. However, for the CCQE background, since the initial and final states are distinct (and nucleons), the electron will tend to scatter at large angles. 
\begin{figure}[!htbp]
\begin{center}
\includegraphics[width=0.75\linewidth]{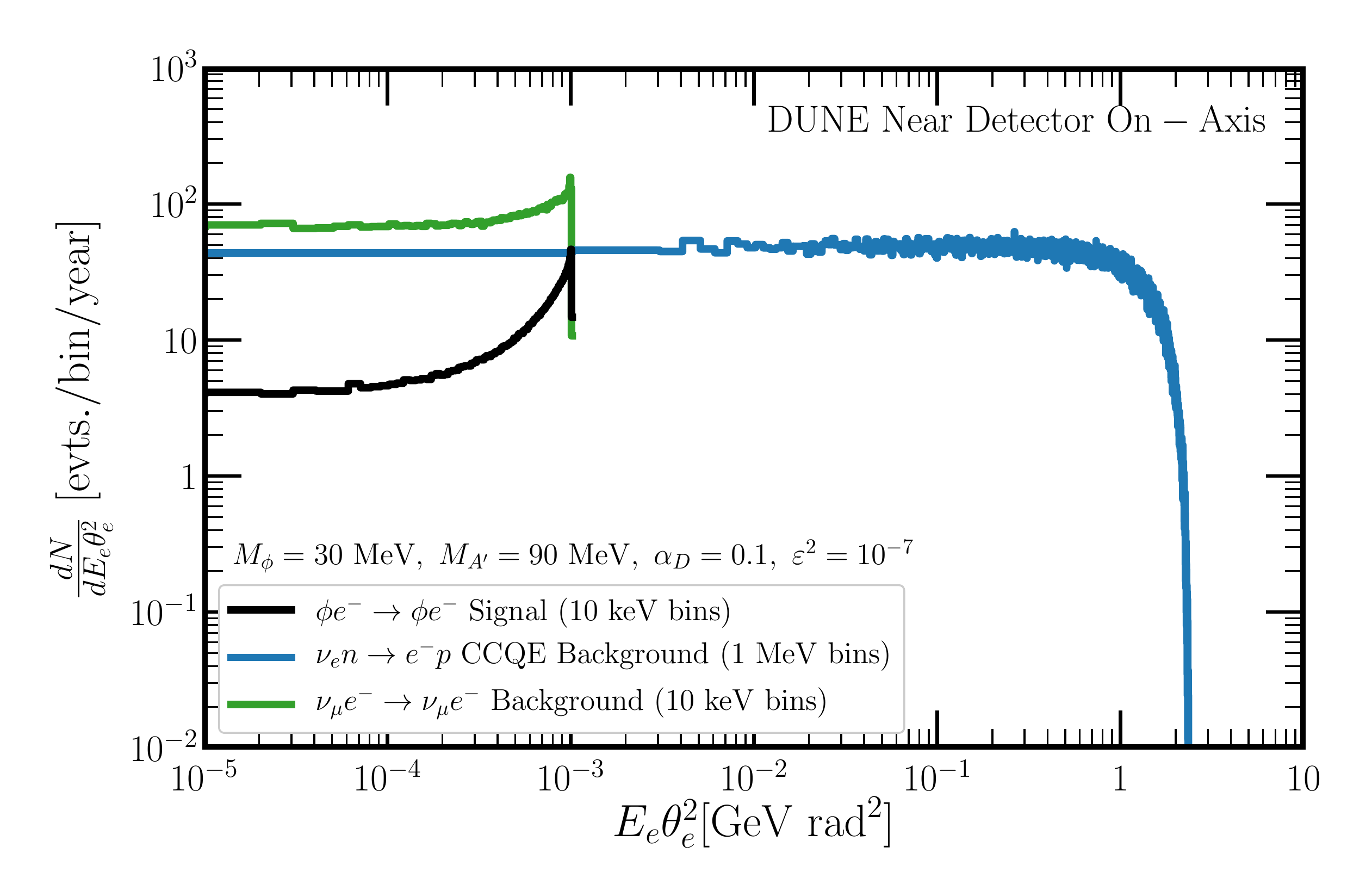}
\caption{Event distributions of $\nu_e$ CCQE background (blue), $\nu$ electron scattering background (green), and $\phi e^-$ signal scattering (black) as a function of $E_e \theta_e^2$, the outgoing electron energy times angle squared. For the signal distribution, we have assumed $M_\phi = 30$ MeV, $M_{A^\prime} = 90$ MeV, $\alpha_D = 0 .1$, and $\varepsilon^2 = 10^{-7}$. Note the different bin widths for the $\nu$ background/signal curves vs. that of the $\nu_e$ CCQE background.}
\label{fig:EThetaSq}
\end{center}
\end{figure}
In Fig.~\ref{fig:EThetaSq} we display the distributions of the signal and backgrounds as a function of $E_e \theta_e^2$ from {\sc MadGraph5}~\cite{Frederix:2018nkq} simulations, integrated over the neutrino (or $\phi$) flux, assuming an on-axis detector.
We find that, for the on-axis detector, fewer than 1 in 1000 CCQE background events have $E_e \theta_e^2 < 2$ MeV rad$^2$ (double the kinematic threshold of the signal and $\nu e^-$ scattering processes), meaning that we can place a cut on this quantity, retaining all of our signal, all of the $\nu e^- \to \nu e^-$ background, and less than $0.1\%$ of the CCQE background. Note that the magnitude of the $\nu_e$ CCQE background has to be rescaled due to the use of different bin widths. A similar analysis performed with the NOvA near detector have shown that it can cut $E_e \theta_e^2 < 5$ MeV rad$^2$~\cite{deNiverville:2018dbu} and remove $\approx 99\%$ of the CCQE background, resulting in the $\nu e^-$ background being the dominant one.  DUNE near detector is expected to have better electron energy and angular resolutions than NOvA, and therefore to further reject the CCQE background.

\section{Sensitivity Improvement from Electron Kinematics}
\label{Appendix:Improvement}

In the previous section, we discussed the capability of vetoing one of the two background channels in the search for dark matter scattering off electrons. Here, we discuss how the sensitivity estimate obtained in may be further improved by including information about the final-state electron kinematics for the signal and background distributions. For simplicity, we focus on the case of the complex scalar DM, but the fermionic DM should provide similar improvement, as their electron scattering cross sections are nearly identical.

In Fig.~\ref{fig:ElectronEnergy} we show the differential event distribution for the background that survives our $E_e \theta_e^2$ cut ($\nu e^- \to \nu e^-$) as well as the signal channel $\phi e^- \to \phi e^-$ for two different choices of $M_\phi$ and $M_{A^\prime}$: in solid lines, $M_\phi = 30$ MeV and $M_{A^\prime} = 90$ MeV; and in dashed lines, $M_\phi = 20$ MeV and $M_{A^\prime} = 30$ MeV. We see that, depending on the DM/$A^\prime$ masses, the electron scattering spectrum can appear significantly different than the background one. This figure is identical to Fig.~\ref{fig:ElectronEnergyMainText} of the main text; we reproduce it here for clarity.
\begin{figure}[!htbp]
\begin{center}
\includegraphics[width=0.75\linewidth]{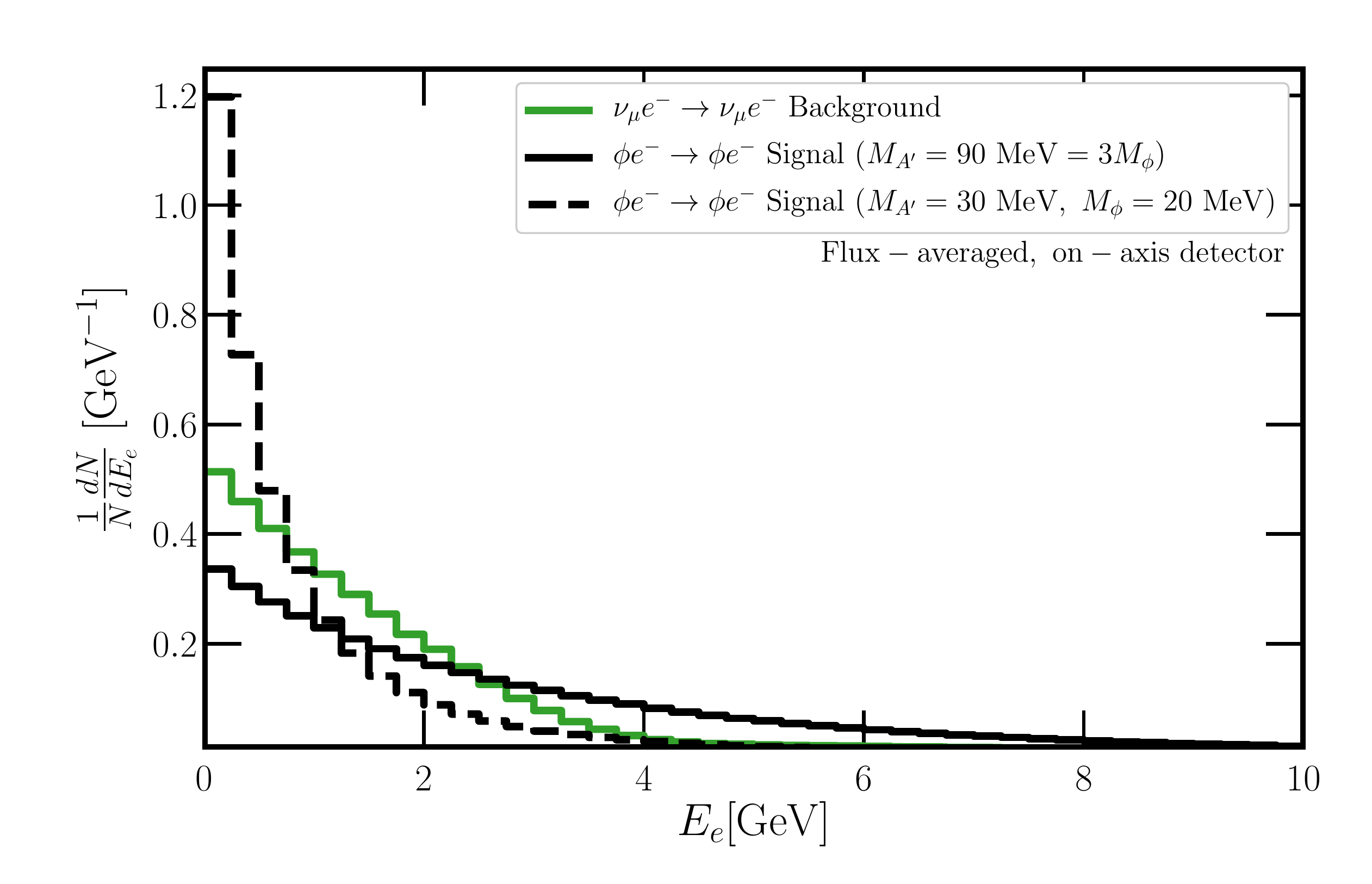}
\caption{(Identical to Fig.~\ref{fig:ElectronEnergyMainText}) Event distributions as a function of electron energy for $\nu e^-$ background (green) and $\phi e^-$ signal (black). The total number of events has been normalized for the two distributions. For the signal distribution, we have assumed $M_\phi = 30$ MeV and $M_{A^\prime} = 90$ MeV for the solid lines, and $M_\phi = 20$ MeV and $M_{A^\prime} = 30$ MeV for the dashed ones.}
\label{fig:ElectronEnergy}
\end{center}
\end{figure}

We will assume a conservative electron energy resolution of $250$ MeV for these events (Ref.~\cite{Acciarri:2015uup} gives the electron energy reconstruction at the level of $\sim 15\%$ for the energies of here, which is much larger than the $250$ MeV bins for the entire $\nu_\mu e^-$ scattering distribution) , and we will consider electron events with energy between $0$ and $10$ GeV (40 bins total). In the main analysis, we were simply performing a counting experiment at each off-axis location, where the number of events expected was at least $\mathcal{O}(100)$ for each location. Now that we are dividing each sample into 40 bins, we will consider Poissonian statistics instead of Gaussian. Instead of an expected number of $\phi$ ($\nu$) events $N_{i}^\phi$ ($N_i^\nu$) at position $i$, we now have an expected number $N_{ij}^{\phi}$ ($N_{ij}^{\nu}$) at position $i$ and electron energy bin $j$. In the absence of uncertainties, our log-likelihood function is
\begin{equation*}
\mathcal{L}_{ij} = -\left( \left(\frac{\varepsilon}{\varepsilon_0}\right)^4 N_{ij}^{\phi} + N^\nu_{ij} \right) + N^\nu_{ij} \ln{\left( \left(\frac{\varepsilon}{\varepsilon_0}\right)^4 N^\phi_{ij} + N^\nu_{ij} \right)} - \ln{\left(N^\nu_{ij}!\right)},
\end{equation*}
and then a sum is performed over $i$ and $j$. Our test statistic is then $-2 \Delta \mathcal{L}$, and we can determine limits on $\varepsilon^2$ accordingly.

We can incorporate a correlated systematic uncertainty on the overall flux as well as uncorrelated systematic uncertainties on each position's normalization, by modifying $\mathcal{L}_{ij}$ as
\begin{equation*}
\mathcal{L}_{ij} \to -A*f_i \left( \left(\frac{\varepsilon}{\varepsilon_0}\right)^4 N^\phi_{ij} + N^\nu_{ij} \right) + N^\nu_{ij} \ln{\left(A*f_i \left( \left(\frac{\varepsilon}{\varepsilon_0}\right)^4 N^\phi_{ij} + N^\nu_{ij} \right)\right)} - \ln{\left(N^\nu_{ij}!\right)},
\end{equation*}
where $f_i$ is the nuisance parameter normalizing the number of events at position $i$ and $A$ is the overall (correlated uncertainty) nuisance parameter. The test statistic then, is
\begin{equation}
-2\Delta \mathcal{L} = \sum_{i = 1} \left[ \sum_{j=1} \left( - 2\mathcal{L}_{ij}\right) + \frac{\left(f_i - 1\right)^2}{\sigma_{f_i}^2}\right] + \frac{\left(A - 1\right)^2}{\sigma_A^2}.
\end{equation}
As in the previous analysis, we assume $\sigma_{f_i} = 1\%$ for all $i$ and $\sigma_A = 10\%$. We marginalize over all $f_i$ and $A$ to calculate our estimated sensitivity. The improvement obtained when considering electron energy leads to roughly a factor of $2$ stronger limits on $\varepsilon^2$ for $A^\prime$ and $\phi$ masses of interest. Note that incorporating this will allow us to probe regions in which the relic abundance of $\chi$ matches the observed dark matter relic abundance for $M_\phi = 20$ MeV, $M_{A^\prime} \simeq 60$ MeV.

\section{Results assuming fermionic Dark Matter}
\label{sec:Fermion}

In this section, we display the results (similar to Figs.~\ref{fig:NSES} and~\ref{fig:2DES} of the main text) assuming that the dark matter is a fermionic particle $\chi$, governed by the Lagrangian

\begin{equation}
\mathcal{L} \supset -\frac{\varepsilon}{2} F^{\mu\nu} F^\prime_{\mu \nu} + \frac{M_{A^\prime}^2}{2} A^\prime_\mu A^{\prime\mu} 
+ \overline{\chi}i \gamma^\mu \left(\partial_\mu - ig_D A^\prime_\mu\right)\chi - M_\chi \overline{\chi}\chi,
\end{equation}


The results assuming $M_{A^\prime} = 3M_{\chi}$ ($M_\chi = 20$ MeV) are shown in Fig.~\ref{fig:FermionLimit} left (right).

The production rate of this DM is very similar to scalar DM, assuming on-shell production. For $M_\chi > M_{A^\prime}/2$ (off-shell production), the fermionic DM production rate is larger than the scalar one, seen in Appendix~\ref{Appendix:Branching}. Additionally, the scattering cross section off electrons is largely similar (Appendix~\ref{Appendix:CrossSections}), so we expect the only difference in sensitivity to be when production is via off-shell $A^\prime$ decays. External constraints (see Fig.~\ref{fig:NSES} of the main text and the discussion in Section~\ref{sec:ExistingLimsDUNE}) will be similar, however smaller values of $\varepsilon$ are required so that the relic abundance of $\chi$ matches the observed relic abundance\footnote{This is because the freeze-out mechanism in the early universe is not p-wave suppressed for fermionic dark matter}. Lastly, precision measurements of the Cosmic Microwave Background (CMB) from the Planck satellite~\cite{Ade:2015xua} set a constraint falling inside the region of interest of our work, shown with an orange dot-dashed line which excludes the region below it. Notice however that this limits holds if $\chi$ is a symmetric thermal relic, while it can be relaxed in scenarios with Majorana or Pseudo-Dirac DM, or if the DM is fermionic and asymmetric. 

\begin{figure*}[!htbp]
\begin{center}
\includegraphics[width=0.8\textwidth]{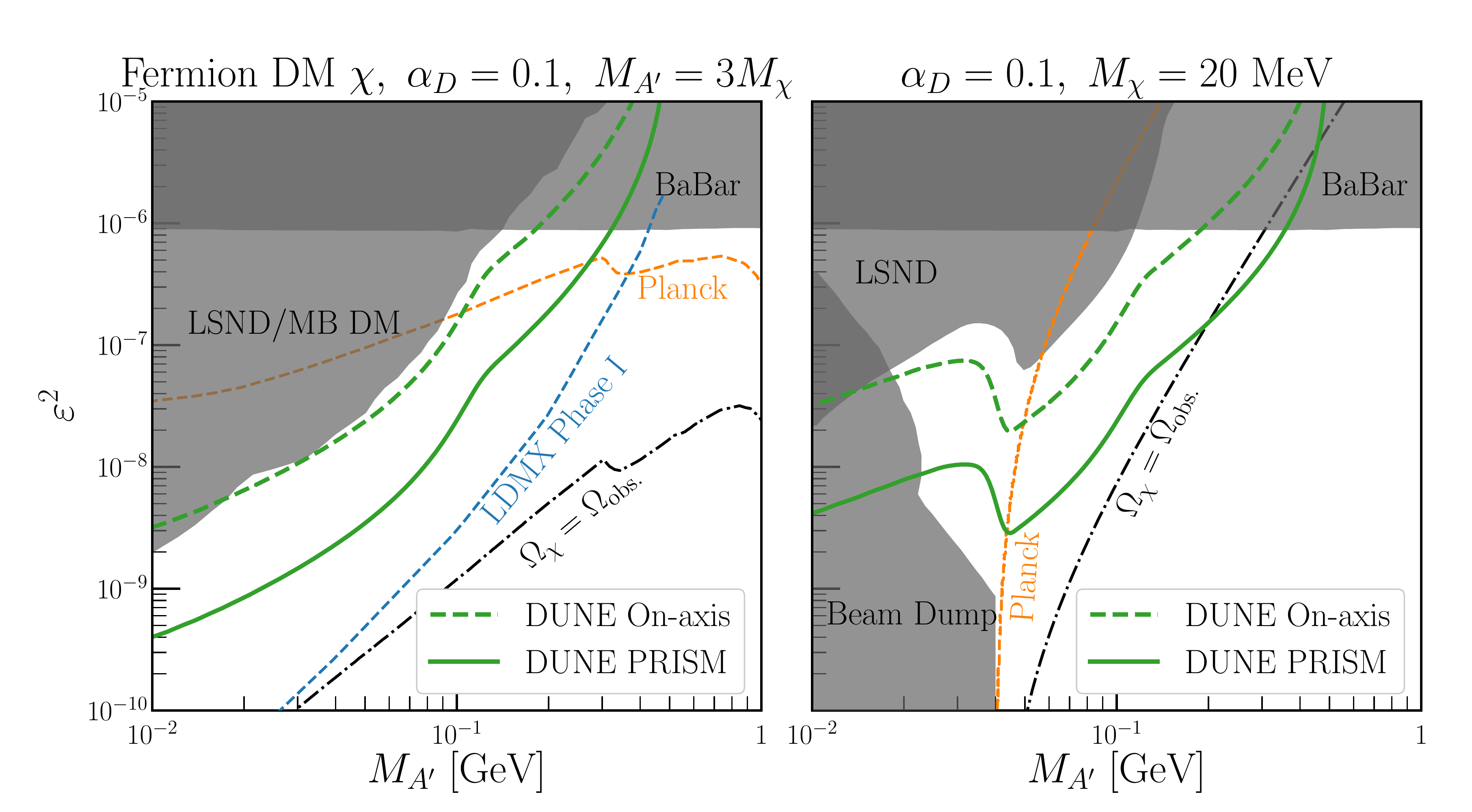}
\caption{Expected DUNE On-axis (dashed) and PRISM (solid) sensitivity using $\chi e^- \to \chi e^-$ scattering. We assume $\alpha_D = 0.1$ in both panels, and $M_{A^\prime} = 3M_\chi$ ($M_\chi = 20$ MeV) in the left (right) panel. Existing constraints are shown in grey, and limits from Planck~\cite{Ade:2015xua} and the relic density target are shown in orange and dot-dashed lines, respectively. We compare our results against the proposed LDMX experiment in blue~\cite{Akesson:2018vlm}.}
\label{fig:FermionLimit}
\end{center}
\end{figure*}

In Fig.~\ref{fig:Fermion2D} we display the results of the two-dimensional scan over $M_{A^\prime}$ and $M_\phi$, showing contours of expected sensitivity for $\varepsilon^2$, fixing $\alpha_D = 0.1$.  In this figure, we also ask the question: for which values of $M_{A^\prime}$ and $M_\chi$ does our limit on $\varepsilon^2$ reach the target for which either the limit on these parameters from Planck is saturated or the relic abundance (assuming $\chi$ is a symmetric thermal relic) target is reached. Those regions are colored in green and grey, respectively.
\begin{figure}[tbp]
\begin{center}
\includegraphics[width=0.5\linewidth]{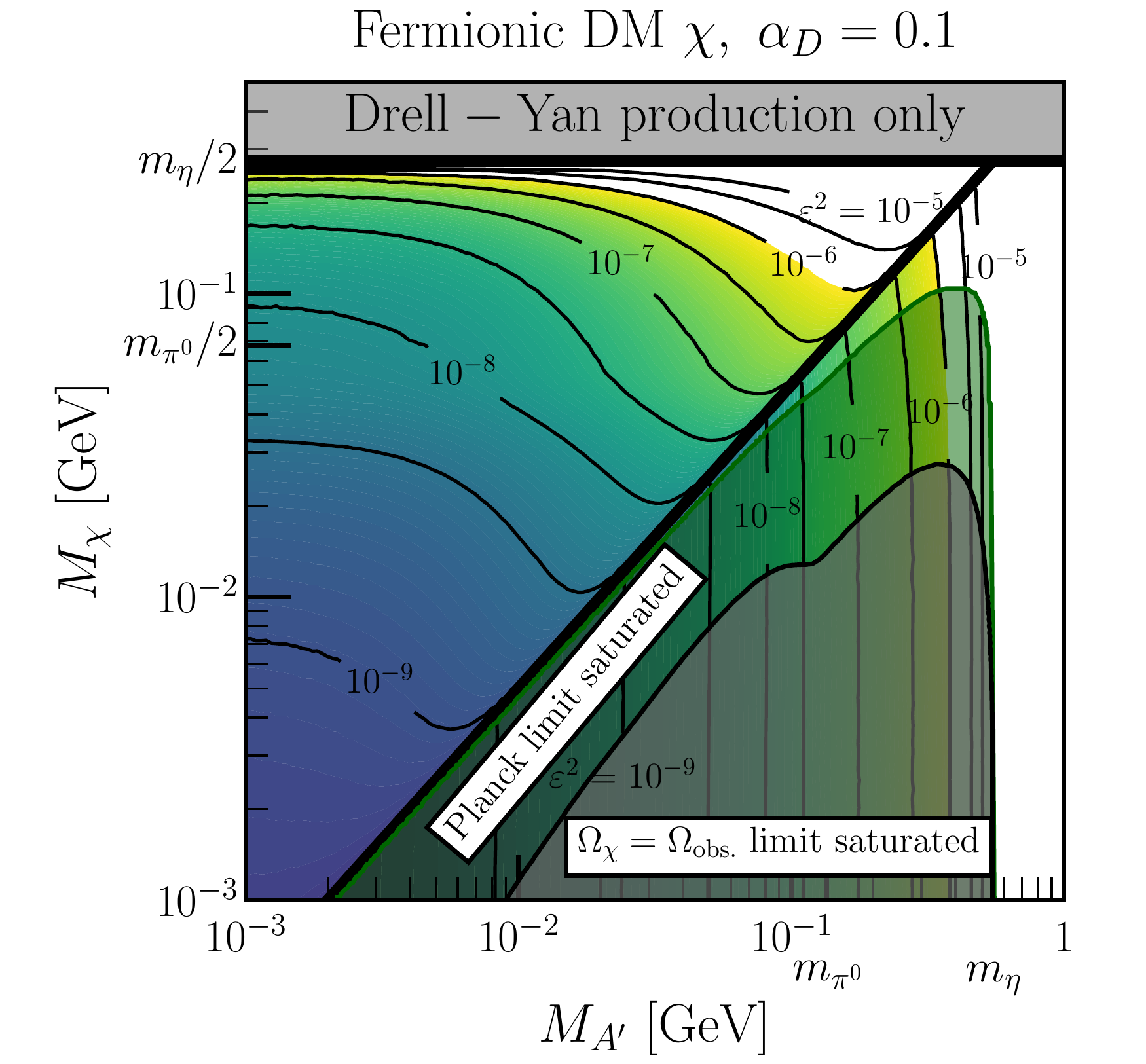}
\caption{Expected limits on $\varepsilon^2$ as a function of $M_{A^\prime}$ and fermion DM mass $M_\chi$ assuming seven years of data collection at DUNE searching for DM scattering off electrons.  We shade in regions for which this expected limit saturates the target for which the DM relic density matches the observed abundance (grey) as well as when it saturates the lower limit from Planck (green).}
\label{fig:Fermion2D}
\end{center}
\end{figure}
\end{widetext}

\bibliography{References}


\end{document}